\documentclass[aps,prapplied,
amsmath,amssymb,reprint]{revtex4-1}

\usepackage{graphicx}
\usepackage{color}
\usepackage{bm}
\usepackage{fancyref}
\usepackage{soul}
\usepackage{float}
\usepackage{tabularx}
\usepackage[T1]{fontenc}       % Use modern font encodings
\usepackage[version=3]{mhchem}
\usepackage{xcolor}

\usepackage{titlesec}
\titlespacing*{\section}{0pt}{1.1\baselineskip}{\baselineskip}

\begin{document}

\author{D. Davidovikj$^1$}
\author{F. Alijani$^2$}
\author{S. J. Cartamil-Bueno$^1$}
\author{H. S. J. van der Zant$^1$}
\author{M. Amabili$^3$}
\author{P. G. Steeneken$^{1,2}$}

\affiliation{$^1$Kavli Institute of Nanoscience, Delft University of Technology, Lorentzweg 1, 2628 CJ Delft, The Netherlands \\
$^2$Department of Precision and Microsystems Engineering, Delft University of Technology, Mekelweg 2, 2628 CD, Delft, The Netherlands \\
$^3$Department of Mechanical Engineering, McGill University, 817 Sherbrooke Street W. Montreal, Quebec, Canada, H3A 2K6}
\title[]{Young's modulus of 2D materials extracted from their nonlinear dynamic response
	}% Force line breaks with \\

\begin{abstract}

	Due to their atomic-scale thickness, the resonances of 2D material membranes show signatures of nonlinearities at amplitudes of only a few nanometers. While the linear dynamics of membranes is well understood, the exact relation between the nonlinear response and the resonator's material properties has remained elusive. In this work, we propose a method to determine the Young's modulus of suspended 2D material membranes from their nonlinear dynamic response. The method is demonstrated by interferometric measurements on graphene and \ce{MoS2} resonators, which are electrostatically driven into the nonlinear regime at multiple driving forces. It is shown that a set of response curves can be fitted by the solutions of the Duffing equation using only one fit parameter, from which the Young's modulus is extracted using membrane theory. Our method is fast, contactless, and provides a platform for high-frequency characterization of the mechanical properties of 2D materials.

\end{abstract}
\maketitle
 The remarkable mechanical properties of 2D material membranes have sparked interest for potential uses as pressure \cite{smith13pressure,dolleman15}, gas \cite{bunch12,dolleman16} and mass \cite{sakhaee08mass,atalaya10} sensors. For such applications it is essential to have accurate methods for determining their mechanical properties. One of the  most striking properties of these ultra-thin materials is their high Young's modulus. In order to measure the Young's modulus, a number of static deflection techniques have been used, including Atomic Force Microscopy (AFM) \cite{hone08elastic,poot08,castellanos12,castellanos12am}, the pressurized blister test~\cite{koenig11} and the electrostatic deflection method~\cite{wong10,nicholl15}. 
The most widely used method is AFM, where by performing a nanoindentation measurement at the center of a suspended membrane, the pre-tension ($n_0$) and Young's modulus ($E$) are extracted from the force-deflection curve. 

Whereas AFM has been the method of choice for static studies, laser interferometry has proven to be an accurate tool for the dynamic characterization of suspended 2D materials, with dynamic displacement resolutions better than 20 fm$/\sqrt{\mathrm{Hz}}$ at room temperature \cite{bunch07,castellanos13,davidovikj16}. Since for very thin structures the resonance frequency is directly linked to the pre-tension in the membrane, these measurements have been used to mechanically characterize 2D materials in the linear limit \cite{bunch07,castellanos13,cartamil15,wang15}. 
At high vibrational amplitudes nonlinear effects start playing a role, which have lately attracted a lot of interest~\cite{eichler2011,croy12,eriksson13,dealba16,mathew16}. In particular, Duffing-type nonlinear responses have been regularly observed \cite{bunch07,chen09,chen13,castellanos13,davidovikj16}. These geometrical nonlinearities, however, have never been related to the intrinsic material properties of the 2D membranes. 

Here, we introduce a method for determining the Young's modulus of 2D materials by fitting their forced nonlinear Duffing response. Using nonlinear membrane theory, we derive an expression that allows us to relate the fit parameters to both the pre-tension and Young's modulus of the material. 
The proposed method offers several advantages: (i) The excitation force is purely electrostatic, requiring no physical contact with the membrane that can influence its shape~\cite{han15,vella2017}; (ii) The on-resonance dynamic operation significantly reduces the required actuation force, compared to static deflection methods; (iii) The high-frequency resonance measurements allow for fast testing by averaging over millions of deflection cycles per second, using mechanical frequencies in the MHz range; (iv) The membrane motion is so fast that slow viscoelastic deformations due to delamination, slippage, and wall adhesion effects are strongly reduced. To demonstrate the method, we measure and analyze the nonlinear dynamic response of suspended 2D nanodrums.

\section{Measurements}

\begin{figure}[ht]
	\includegraphics[width=0.45\textwidth]{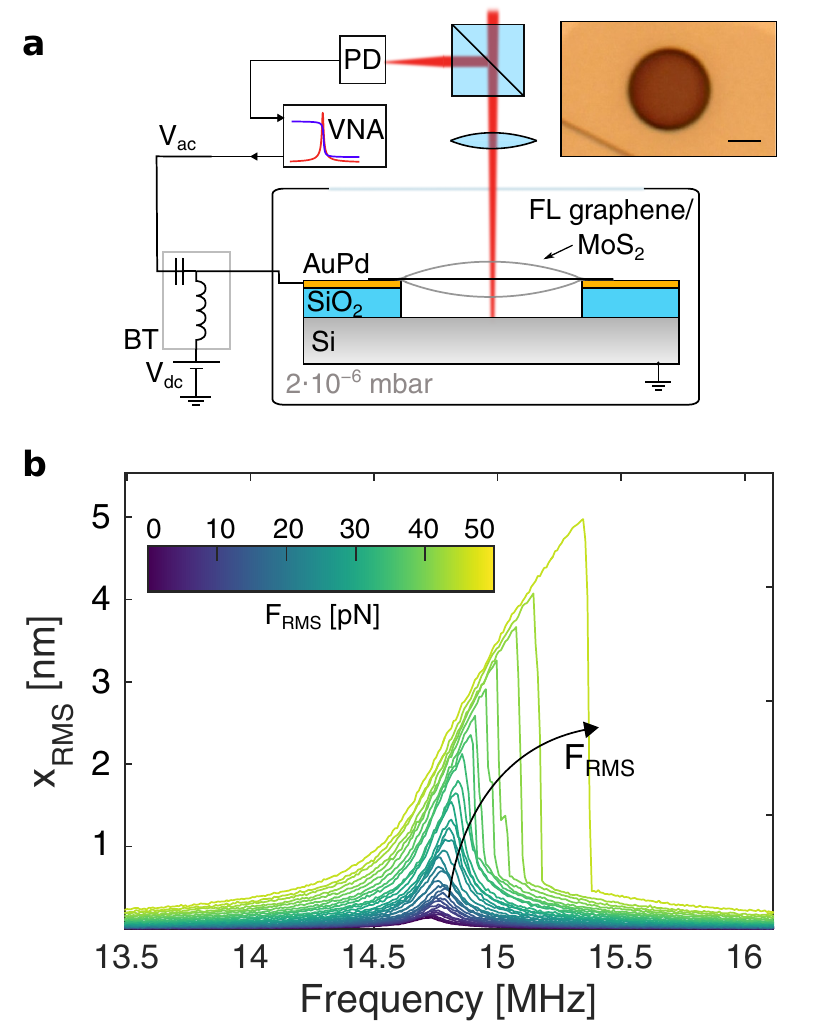}
	\caption{\label{fig:Fig1} (a) Schematic of the measurement setup: a laser interferometer setup is used to read out the motion of the nanodrum. The Si substrate is grounded and, using a bias-tee (BT), a combination of ac- and dc voltage is applied to electrostatically actuate the motion of the drum. This motion modulates the reflected laser intensity and the modulation is read out by a photodiode. Inset: an optical image of a FL graphene nanodrum (scale bar: 2 $\mathrm{\mu m}$). (b) Frequency response curves of the calibrated root-mean-square (RMS) motion amplitude for increasing electrostatic driving force. The onset of nonlinearity is visible above $F_\mathrm{RMS}$ = 15 pN. The color of the curves indicates the corresponding driving force.}
\end{figure}

The samples consist of cavities on top of which exfoliated flakes of 2D materials are transferred using a dry transfer technique \cite{castellanos14}. One of the measured devices, a few-layer (FL) graphene nanodrum, is shown in the inset of Fig.~\ref{fig:Fig1}(a). The measurements are performed in vacuum at room temperature. Electrostatic force is used to actuate the membrane and a laser interferometer to detect its motion, as described in \cite{bunch07,castellanos13,wang15,davidovikj16}. A schematic of the measurement setup is shown in Fig.~\ref{fig:Fig1}(a).~The details on the sample preparation and measurement setup are described in the Experimental Section below. 

Fig. \ref{fig:Fig1}(b) shows a set of calibrated frequency response curves of the fundamental mode of this graphene drum (with thickness $h = 5 $ nm and radius $R = 2.5\,\mathrm{\mu m}$) driven at different ac voltages ($V_\mathrm{ac}$). The dc voltage is kept constant ($V_\mathrm{dc} = 3$ V) throughout the entire measurement with $V_\mathrm{dc}\gg V_\mathrm{ac}$. All measurements are taken using upward frequency sweeps. The RMS force $F_\mathrm{RMS}$ is the root-mean-square of the electrostatic driving force. For high driving amplitudes ($F_\mathrm{RMS} > 15 \mathrm{pN}$), the resonance peak starts to show a nonlinear hardening behavior, which contains information on the cubic spring constant of the membrane.

\section{Fitting the nonlinear response}

We can approximate the nonlinear response of the fundamental resonance mode by the Duffing equation (see Section I of the Supporting Information):

\begin{equation}
m_\mathrm{eff}\ddot{x} + c\dot{x} + k_1 x + k_3 x^3 = \xi F_\mathrm{el} cos(\omega t),
\label{Duffing}
\end{equation}

\noindent where $x$ is the deflection of the membrane's center, $c$ is the damping constant, $k_1$ and $k_3$ are the linear and cubic spring constants and $m_\mathrm{eff} = \alpha m$ and $\xi F_\mathrm{el}$ are the mass and the applied electrostatic force corrected by factors ($\alpha$ and $\xi$) that account for the mode-shape of the resonance (for a rigid-body vertical motion of the membrane $\alpha$ and $\xi$ are both 1). As shown in the Supporting Information Section I, for the fundamental mode of a fixed circular membrane $\xi=0.432$ and $\alpha=0.269$. The parameters in the Duffing equation (\ref{Duffing}) are related to the resonance frequency $\omega_0$ ($\omega_0 = 2\pi f_0$) and the $Q$-factor by $Q=\omega_0 m_\mathrm{eff}/c$ and $\omega_0^2=k_1/m_\mathrm{eff}$. 

The fundamental resonance frequency ($f_0 = 14.7$ MHz) is extracted from the linear response curves at low driving powers (Fig.~\ref{fig:Fig1}(b)), and is directly related to the pre-tension ($n_0$) of the membrane: $n_0 = 0.69 \pi^2 f_0^2 R^2\rho h $, where $\rho$ is the mass density of the membrane (in this case $n_0 =$ 0.107 N/m). In order to fit the set of nonlinear response curves, the steady-state solution of the Duffing equation (eq. \ref{Duffing}) is converted to a set of algebraic equations using the harmonic balance method (see Section II of the Supporting Information). Using these equations, the entire set of curves can then be fitted by a least-squares optimization algorithm. Since $N$ curves are fitted simultaneously, the expected fitting error is roughly a factor $\sqrt{N}$ lower than that of single curve fit.

\begin{figure}[ht]
	\includegraphics[width=0.45\textwidth]{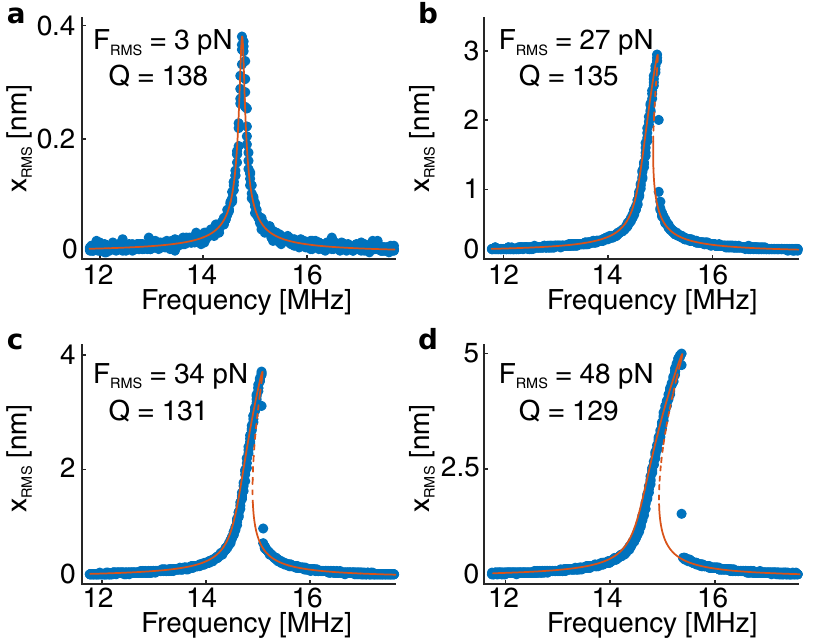}
	\caption{\label{fig:Fig2} Measured traces (blue scatter plot) and the corresponding fits (red curves) showing both the stable (solid line) and the unstable (dashed line) solutions of the Duffing equation. (a)-(d) are frequency response curves of the device from Fig.~\ref{fig:Fig1} at four different driving forces, denoted in the top left corner of each panel, along with the extracted $Q$-factors. The extracted cubic spring constant is $k_3$ = 1.35 $\cdot 10^{15} \mathrm{N/m^3}$.}
\end{figure}
\bigskip
The $Q$-factor is implicitly related to $k_3$ by a function $Q_\mathrm{i}=Q_\mathrm{i}(k_3,A_\mathrm{max,i},F_\mathrm{el,i})$, where $A_\mathrm{max,i}$ are the peak amplitudes and $F_\mathrm{el,i}$ are the driving force amplitudes for each of the measured curves~\cite{lifshitz08,amabili16} (see Section II of the Supporting Information). The amplitudes $A_\mathrm{max,i}$ are found from the experimental data and the whole dataset is fitted using a single fit parameter: the cubic spring constant $k_3$. The results of this procedure are presented in Fig.~\ref{fig:Fig2}(a-d), which shows four frequency response curves and their corresponding fits. The solutions of the steady-state amplitude for the Duffing equation (red curves in Fig.~\ref{fig:Fig2}) are plotted by finding the positive real roots $x^2$ of: 
\begin{eqnarray}
\xi^2 F_\mathrm{el}^2&=& (\omega^2 c^2 + m_\mathrm{eff}^2(\omega^2-\omega_0^2)^2) x^2 \nonumber \\
& &- \frac{3}{2} m_\mathrm{eff}(\omega^2-\omega_0^2) k_3 x^4 + \frac{9}{16} k_3^2 x^6.
\label{Duffing2}
\end{eqnarray}

A good agreement between fits and data is found using the single extracted value $k_3$~=~$1.35~\cdot~10^{15}~ \mathrm{N/m^3}$, which demonstrates the correspondence between the measurement and the underlying physics. We note that at higher driving amplitudes, we also observe a reduction in the $Q$-factor (by nearly 10\% at the highest measured driving amplitude). This can be a signature of nonlinear damping mechanisms which is in line with previously reported measurements on graphene mechanical resonators  ~\cite{eichler2011,croy12,singh16}. In the following section, we will lay out the theoretical framework to relate the extracted cubic spring constant $k_3$ to the Young's modulus of the membrane.

\section{Theory}

The nonlinear mechanics of a membrane can be related to its material parameters via its potential energy. The potential energy of a radially deformed circular membrane with isotropic mechanical properties can be approximated by a function of the form:
\begin{equation}
U = \frac{1}{2} C_1(\nu)n_0 x^2 + \frac{1}{4} C_3 (\nu)\frac{E h\pi}{R^2} x^4,
\label{potE}
\end{equation}
where $R$ and $h$ are the membrane's radius and thickness respectively. Bending rigidity is neglected, which is a good approximation for $h/R > 1/1000$~\cite{mansfield2005}. $C_1(\nu)$ and $C_3(\nu)$ are dimensionless functions that depend on the deformed shape of the membrane and the Poisson's ratio $\nu$ of the material. The term in equation (\ref{potE}) involving $C_1$ represents the energy required to stretch a membrane under a constant tensile pre-stress, the $C_3$ term signifies that the tension itself starts to increase for large membrane deformations. The out-of-plane modeshape for the fundamental resonance mode of a circular membrane is described by a zero-order Bessel function of the first kind ($J_0(r)$). Numerical calculations of the potential energy (\ref{potE}) of this mode give $C_1(\nu)=1.56\pi n_0$ and $C_3(\nu)=1/(1.269-0.967 \nu - 0.269 \nu^2)$ (see Section I of the Supporting Information). Using equation (\ref{potE}) the nonlinear force-deflection relation of circular membranes is given by

\begin{equation}
\label{forceDefl}
F= \frac{dU}{dx} = k_1 x + k_3 x^3 =  C_1(\nu)n_0 x + C_3 (\nu)\frac{E h\pi}{R^2}x^3.
\end{equation}

The functions $C_1$ and $C_3$ have previously been determined for the potential energies of statically deformed membranes by AFM ~\cite{komaragiri05,castellanos12} and uniform gas pressure ~\cite{hencky15,boddeti13}. Their functional dependence depends entirely on the shape of the deformation of the membrane. In Table 1 we summarize the functional dependences of $k_1$ and $k_3$ for the 3 types of membrane deformation.

By combining eq. ~\ref{forceDefl} with the obtained functions for $C_1$ and $C_3$ from Table 1 (last row), the Young's modulus $E$ can be determined from the cubic spring constant $k_3$ by
\begin{equation}
E = \frac{(1.27-0.97 \nu - 0.27 \nu^2)R^2}{\pi h}k_3.
\end{equation}

	\renewcommand{\arraystretch}{2}%
		\begin{tabular}{ |c|c|c|c| } 
			
			\cline{2-4}
			\multicolumn{1}{c|}{}& $k_1$ & $k_3$ & Def. shape\\ 
			\hline
			AFM & $ \pi n_0$ & $\frac{1}{(1.05-0.15\nu-0.16\nu^2)^3}\frac{E h}{R^2}$ & \includegraphics[scale=0.8]{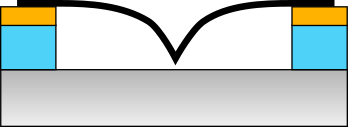}\\ 
			\hline
			$\Delta P$ & $4\pi n_0$ & $\frac{8\pi}{3(1-\nu)}\frac{E h}{R^2}$ & \includegraphics[scale=0.8]{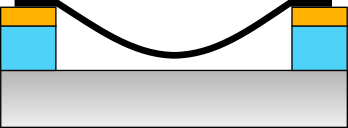}\\ 
			\hline
			This work & $1.56\pi n_0$ & $\frac{\pi}{1.27-0.97 \nu - 0.27 \nu^2}\frac{ E h}{R^2}$ & \includegraphics[scale=0.8]{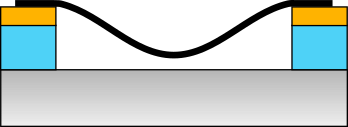}\\ 
			\hline
			
		\end{tabular}
\bigskip

\small \noindent Table 1. $k_1$ and $k_3$ for AFM nanoindentation (AFM), bulge testing of membranes ($\Delta P$) and the nonlinear dynamics method (this work) for the fundamental resonance mode. The corresponding deformation shape, which determines the functional dependence of $k_1$ and $k_3$, is shown on the right.
\bigskip

\begin{figure}[ht]
	\includegraphics[width=0.45\textwidth]{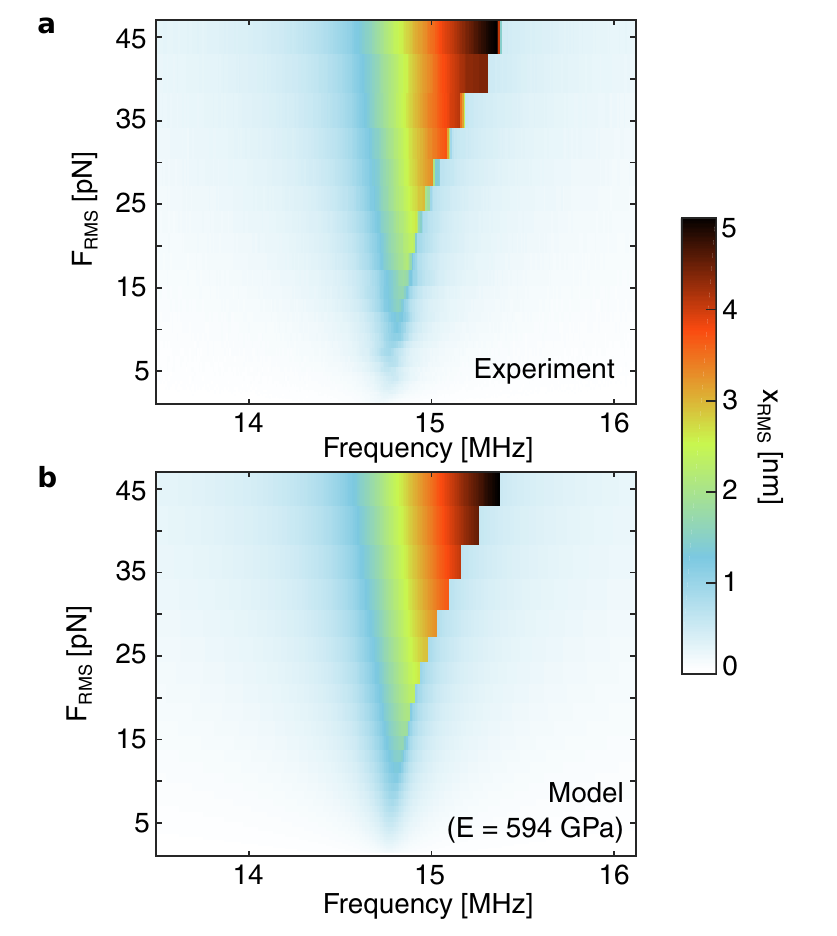}
	\caption{\label{fig:Fig4} Comparison of the RMS motion amplitude ($x_\mathrm{RMS}$) between experiment (a) and model (b) using the identified value for the Young's modulus ($E$ = 594 GPa) for the device shown in Fig.~\ref{fig:Fig1}. }
\end{figure}

From this equation, with the value of $k_3$ extracted from the fits, a Young's modulus of $E$~=~$594~\pm~45$~GPa is found, which is in accordance with literature values which range from  $430 - 1200$ GPa ~\cite{castellanos15_review,isacsson2017_review}. Using this value, the nonlinear dynamic response of the system can be modeled for different driving powers and frequencies. Figure~\ref{fig:Fig4} shows color plots representing the RMS amplitude of the motion of the membrane center as a function of frequency and driving force. Excellent agreement is found between the experiment (Fig. ~\ref{fig:Fig4}(a)) and the model (Fig. ~\ref{fig:Fig4}(b)). 

\begin{figure}[ht]
	
	\includegraphics[width=0.4\textwidth]{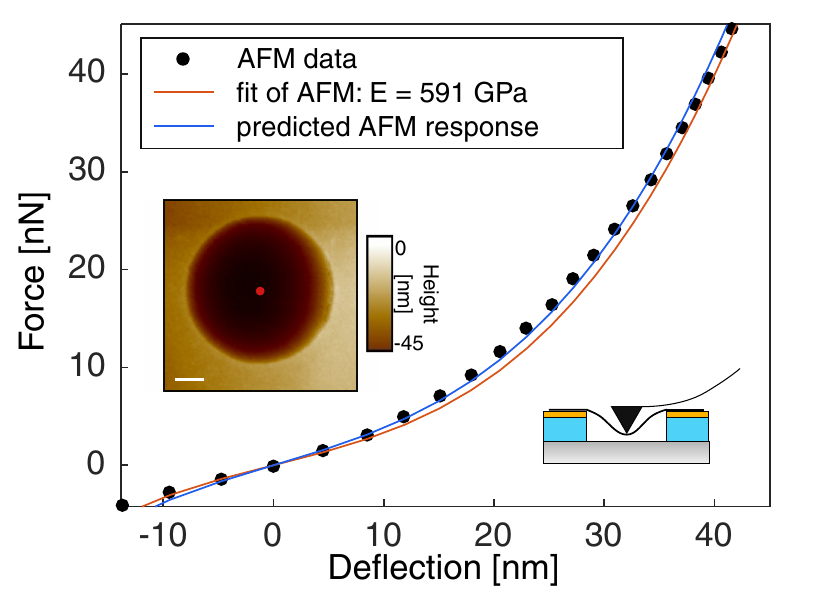}
	\caption{\label{fig:afm}AFM force-deflection curve during tip retraction and the corresponding fit (red curve). Inset shows the AFM image of the drum (scale bar is 1 $\mathrm{\mu m}$). The curve is taken at the center of the drum from Fig.~\ref{fig:Fig1} (marked by the red dot in inset). The blue curve represents the predicted AFM response using the $n_0 = 0.107$ N/m and $E=594$ GPa, obtained from the fit of the nonlinear dynamic response.}
	
\end{figure}

In order to confirm the validity of the method, we performed an AFM nanoindentation measurement on the same graphene drum. A force-deflection measurement, taken at the center of the drum, is plotted in Fig.~\ref{fig:afm} (black dots). The curve is fitted by the AFM force-deflection equation given in Table 1, yielding $E$=591 GPa and $n_0$=0.093 N/m (red curve in Fig.~\ref{fig:afm}). The blue curve shows the expected force-deflection curve based on the values for the Young's modulus and pre-tension extracted from the nonlinear dynamic response fits. The two curves are in close agreement.

\begin{figure}[ht]
	\includegraphics[width=0.33\textwidth]{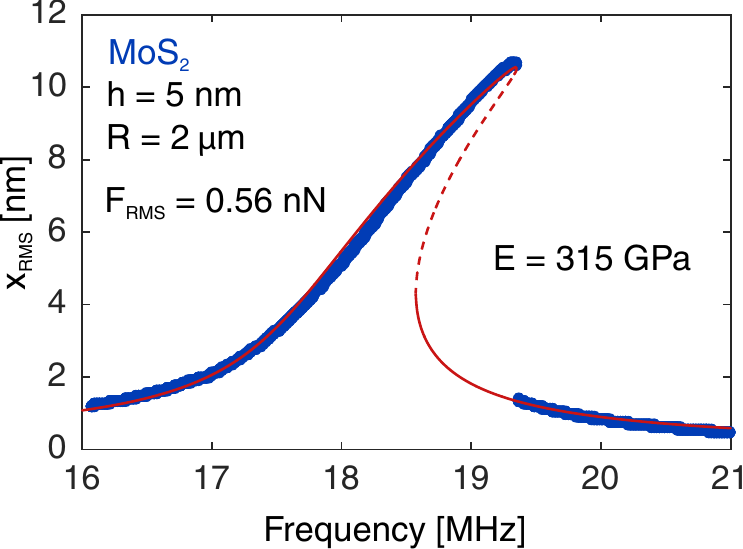}
	\caption{\label{fig:Fig5} Measurement (blue dots) and fit (drawn red curve: stable solutions; dashed red curve: unstable solutions) of a 5 nm thick \ce{MoS2} drum with a Young's modulus of 315 GPa. }

\end{figure}

Finally, to demonstrate the versatility of the method, additional measurements on an \ce{MoS2} nanodrum are presented in Fig. ~\ref{fig:Fig5}. The extracted Young's modulus of \ce{MoS2} ($E = 315 \pm 23$ GPa) is also in agreement with literature values ($E_\mathrm{MoS_2} = 140 - 430$ GPa \cite{castellanos12,castellanos15_review}). The extracted pre-tension of the \ce{MoS2} drum is $n_0 = 0.22$ N/m.

\section{Discussion}

There are several considerations that one needs to be aware of when applying the proposed method. In an optical detection scheme, as the one presented in this work, the cavity depth has to be optimized so that the photodiode voltage is still linearly related to the motion at high amplitudes and the power of the readout laser has to be kept low to avoid significant effects of optothermal back-action~\cite{barton12}. The proposed mathematical model assumes that the bending energy is much smaller than the membrane energy. This is valid for membranes under tension (thickness-to-radius ratio $h/R<0.001$) \cite{mansfield2005}, as is most often the case with suspended 2D materials \cite{bunch07,castellanos13,cartamil15}. It is noted that the electrostatic force also has a nonlinear spring-softening component due to its displacement amplitude dependence. However, in the current study, the vibration amplitudes are much smaller than the cavity depth and this contribution can be safely neglected (see Section III of the Supporting Information for derivation).

Compared to conventional mechanical characterization methods~\cite{hone08elastic,poot08,castellanos12,castellanos12am,koenig11,wong10,nicholl15}, the presented method provides several advantages. Firstly, no physical contact to the flake is required. This prevents effects such as adhesion and condensation of liquids between an AFM tip and the membrane, that can influence the measurements. Moreover, the risk of damaging the membrane is significantly reduced. The on-resonance operation allows the usage of very small actuation forces, since the motion amplitude at resonance is enhanced by the $Q$-factor. Unlike AFM, where the force is concentrated in one point, here the force is more equally distributed across the membrane, resulting in a more uniform stress distribution. Additionally, for resonators with a high quality factor, the modeshape of vibrations is practically independent of the shape or geometry of the actuator.

The high-frequency nature of the presented technique is advantageous, since it allows for fast characterization of samples, and might even be extended for fast wafer-scale characterization of devices. Every point of the frequency response curve corresponds to many averages of the full force-deflection curve (positive and negative part) which reduces the error of the measurement and eliminates the need of offset calibration of the zero point of displacement~\cite{lifshitz08}. The close agreement between the AFM and nonlinear dynamics value for the Young's modulus $E$ indicates that viscoelasticity, and other time dependent effects like slippage and relaxation, are small in graphene. Therefore, the dynamic stiffness is practically coinciding with the static stiffness. For future studies it is of interest to apply the method to study viscoelastic effects in 2D materials, where larger differences between AFM and resonant characterization measurements are expected.

\section{Conclusion}

In conclusion, we provide a contactless method for characterizing the mechanical properties of suspended 2D materials using their nonlinear dynamic response. A set of nonlinear response curves is fitted using only one fit parameter: the cubic spring constant. Mathematical analysis of the membrane mechanics is used to relate the Duffing response of the membrane to its material and geometrical properties. These equations are used to extract the pre-tension and Young's modulus of both graphene and MoS$_2$, which are in close agreement with nanoindentation experiments. The non-contact, on-resonant, high-frequency nature of the method provides numerous advantages, and makes it a powerful alternative to AFM for characterizing the mechanical properties of 2D materials. We envision applications in metrology tools for fast and non-contact characterization of 2D membranes in commercial sensors and actuators.\\

\section{Experimental section}

\textit{Sample fabrication.}
A chip with cavities is fabricated from a thermally oxidized Si wafer, with a \ce{SiO2} thickness of 285 nm, using standard lithographic and metal deposition techniques. Circular cavities are etched into the oxide by using a 100 nm gold-palladium ($\mathrm{Au_{0.6}Pd_{0.4}}$) hard mask, which also functions as an electrical contact to the 2D flake. The final depth of the cavities is $g=$385 nm and their radii are $R = 2 - 2.5\mathrm{\mu m}$. The flakes of graphene and \ce{MoS2} are exfoliated from natural crystals.\\
\textit{Measurement setup.}
The sample is mounted in a vacuum chamber ($2\cdot 10^{-6}$ mbar) to minimize damping by the surrounding gas. Using the silicon wafer as a backgate, the membrane is driven by electrostatic force and its dynamic motion is detected using a laser interferometer (see ~\cite{davidovikj16}). The detection is performed at the center of the drum, using a Vector Network Analyzer (VNA). A dc voltage ($V_\mathrm{dc}$) is superimposed on the ac output of the VNA ($V_\mathrm{ac}$) through a bias-tee (BT), such that the small-amplitude driving force at frequency $\omega$ is given by
$F_\mathrm{el}(t) =~\varepsilon_0 R^2\pi V_\mathrm{dc}V_\mathrm{ac}\cos{(\omega t)}/d^2$. The measured VNA signal (in V/V) is converted to a root-mean-squared amplitude ($ x _\mathrm{RMS}$) of the drum motion, using a calibration measurement of the thermal motion taken with a spectrum analyzer \cite{bunch07,hauer13,davidovikj16}.

\section{Acknowledgments}

This work was supported by the Netherlands Organisation for Scientific Research (NWO/OCW), as part of the Frontiers of Nanoscience (NanoFront) program and the European Union Seventh Framework Programme under grant agreement $\mathrm{n{^\circ}~604391}$ Graphene Flagship.\\

\pagebreak

\onecolumngrid
\appendix
\newpage
\section*{Supporting Information}
\setcounter{figure}{0}  
\subsection*{1. Equations of motion}
The strain energy of the circular membrane can be obtained as~\cite{amabili08}
\begin{equation}
U=\int_{0}^{2\pi}\int_{0}^{R} \frac{ Eh}{2 (1-\nu^2)} \Big(\epsilon_{rr}^2+\epsilon_{\theta \theta}^2+2\nu \epsilon_{rr} \epsilon_{\theta \theta}+\frac{1-\nu}{2}\gamma_{r \theta}^2\Big) r dr d\theta ,
\end{equation}
where $E$ is the Young's modulus, $\nu$ is the Poisson's ratio,  $h$ is the thickness and $R$ is the radius of the membrane. Moreover,  $\epsilon_{rr}$, $\epsilon_{\theta \theta}$, and 
$\gamma_{r \theta}$ are the normal and shear strains that are determined as

\begin {equation}
\epsilon_{rr}=\frac{\partial u}{\partial r}+\frac{1}{2}\Big(\frac{\partial w}{\partial r}\Big)^2 ,
\end{equation}
\begin {equation}
\epsilon_{\theta \theta}=\frac{\partial v}{r \partial \theta}+\frac{u}{r}+\frac{1}{2}\Big(\frac{\partial w}{r \partial \theta}\Big)^2,
\end{equation}
\begin {equation}
\gamma_{r \theta}=\frac{\partial v}{ \partial r}-\frac{v}{r}+\frac{\partial u}{r \partial \theta}+\Big(\frac{\partial w}{\partial r}\Big)\Big(\frac{\partial w}{r \partial \theta}\Big),
\end{equation}

where $u$, $v$ and $w$ are the radial, tangential and transverse displacements respectively. For a membrane with fixed edges $u$ and $w$ shall vanish at $r=R$. Moreover, $u$ 
should be zero at $r=0$ for continuity and symmetry. Assuming only axisymmetric vibrations ($v=0$ and $\partial /\partial \theta = 0$) and fixed edges, the solution is approximated as~\cite{timoshenko59}

\begin{subequations}
	\begin{equation}
	w=x(t) J_{0} \Big(\alpha_{0}\frac{r}{R}\Big) , 
	\end{equation}
	\begin{equation}
	u= {u_{0} r}+ r (R-r) \sum_{k=1}^{\bar{N}} q_{k}(t) r^{k-1}.
	\end{equation}
\end{subequations}

Here it should be noted that for axisymmetric vibrations the shear strain $\gamma_{r \theta}$ would become zero. In eqs. (5a,b), $x(t)$ is the generalized coordinate associated with the fundamental axisymmetric mode and $q_{k}(t)$ are the generalized coordinates associated with the radial motion.
Moreover, $J_{0}$ is the Bessel function of order zero, and $\alpha_{0}=2.40483$. 
In addition, $\bar{N}$ is the number of necessary terms in the expansion of radial displacement, and $u_{0}$ is the initial displacement due to pre-tension $n_{0}$ that is obtained 
from the initial stress $\sigma_0=n_{0}/h$ as follows:
\begin {equation}
u_{0}=\frac {\sigma_{0} (1-\nu)}{E}.
\end{equation}

The kinetic energy of the memebrane neglecting radial (i.e. in-plane) inertia , is given by
\begin{equation}
T=\frac{1}{2}\rho h\int_{0}^{2\pi} \int_{0}^{R} \dot{w^2} r dr d\theta,
\end{equation}
where the overdot indicates differentiation with respect to time $t$. 
\par In the presence of transverse harmonic distributed force of constant direction, the virtual work done is 
\begin {equation}
W=2\pi\int_{0}^{R} p w r dr=\frac{2} {R^2}\int_{0}^{R} F_\mathrm{el} cos(\omega t) w r dr,
\end{equation}

where $\omega$ is the excitation frequency and $F_{el}$ gives the force amplitude, positive in transverse direction. Higher-order terms in w are neglected in eq(8) ~\cite{amabili15}. The Lagrange equations of motion are
\begin{equation}
\frac {d}{dt} (\frac{\partial T}{\partial \dot{\textbf{q}}})-\frac{\partial T}{\partial \textbf{q}}+\frac{\partial U}{\partial \textbf{q}}=\frac{\partial W}{\partial \textbf{q}} ,
\end{equation}
and $\textbf{q}=[x(t),q_k(t)] , k=1,...,\bar{N}$ , is the vector including all the generalized coordinates.

Since radial inertia has been neglected, eq. (9) leads to a system of nonlinear equations comprising of a single differential equation associated with the generalized coordinate $x(t)$ and $\bar{N}$ algebraic equations in terms of $q_{k}(t)$. By solving the $\bar{N}$ algebraic equations it is  possible to determine $q_{k}(t)$ 
in terms of $x(t)$. This will reduce the set of nonlinear equations to a single Duffing oscillator as follows:

\begin{equation}
m_\mathrm{eff} \ddot{x}+c \dot{x}+k_{1}x+k_{3}x^3=\xi F_{el} cos(\omega t),
\end{equation}
where 
\begin{equation}
m_\mathrm{eff}=0.847\rho hR^2  , \,   k_{1}=4.897 n_{0}  , \,    \xi=0.432,
\end{equation}

\noindent and $c$ is the damping coeficient that has been added to the equation of motion to introduce linear viscous dissipation. Moreover, $k_{3}$ is the cubic stiffness, which is a function of the Young's modulus and the Poisson's ratio, and its convergence and accuracy is determined by using different number of terms in the expression of the radial displacement (eq. (5b)). The value of $k_3$ converges for $\bar{N}>3$ and its relation to the Young's modulus can be determined by fixing the value of the Poisson's ratio and numerically solving the set of $\bar{N}$ Lagrange equations. $k_3$ can be expressed in the form:

\begin{equation}
k_3 = C_3 (\nu) \frac{Eh\pi}{R^2},
\end{equation}

\noindent where $C_3$ is dimensionless constant which is a function of the Poisson's ratio. The solutions for $C_3$ as a function of $\nu$ are plotted in Figure S1 for values of  the Poisson's ratio between 0 and 0.35.

\begin{figure}[ht]
	\includegraphics[width=0.45\textwidth]{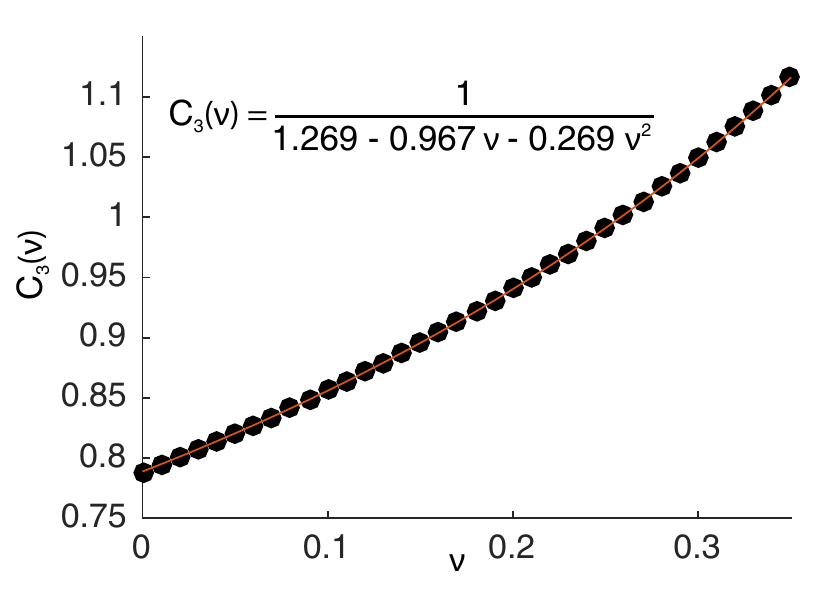}
	\caption{\label{fig:FigS1} Numerical solutions for $C_3$ as a function of $\nu$ and the corresponding fit (red line).  }
\end{figure}

The relation between $C_3$ and $\nu$ is best described with a second-order polynomial, namely:

\begin{equation}
C_3 = \frac{1}{1.269-0.967 \nu - 0.269 \nu^2}.
\end{equation}

This functional dependence is similar to the one used for AFM nanoindentation measurements, often referred to as $q(\nu)$ ~\cite{hone08elastic}.

Next, the following dimensionless parameters are introduced:
\begin{subequations}
	\begin{equation}
	\hat{t}=\omega t ,
	\end{equation}
	\begin{equation}
	\hat{x}=x/h.
	\end{equation}
\end{subequations}
By using eqs. (10) and (14) the following dimensionless equation of motion can be obtained:

\begin{equation}
r^2 \ddot{\hat{x}}+\frac {1} {Q} r \dot{\hat{x}}+\hat{x}+\eta_{3}\hat{x}^3=\lambda cos(t) ,
\end{equation}
where
\begin{equation}
\omega_{0}=\sqrt{\frac{k_{1}}{m}} , Q=\frac{m\omega_{0}}{c} , \eta_3=\frac{k_{3} h^2}{k_{1}} , \lambda=\frac{\xi F_{el}}{m\omega_{0}^2 h} , r=\frac{\omega}{\omega_{0}} .
\end{equation}
Eq. (15) is valid for studying nonlinear vibrations of membranes subjected to external harmonic excitation in the frequency neighborhood of the fundamental mode if the fundamental mode 
of vibration is not involved in an internal resonance with other modes. If such condition retains, then other modes accidentally excited will decay with time to zero due to the presence of damping~\cite{nayfeh08}.
In this work, it is assumed that this condition is preserved and therefore the response of the membrane is described by a single Duffing oscillator for performing nonlinear parameter estimation.

\subsection*{2. Nonlinear identification}
In order to obtain the coefficients of the Duffing oscillator, here we utilize the harmonic balance method. This method is a suitable and accurate mathematical technique that entails the solution of nonlinear equations to be approximated by a truncated Fourier series.
In case of  the dimensionless Duffing oscillator (i.e. eq. (15)), a first order trucation has been shown to provide accurate results~\cite{nayfeh08}. Hence,
\begin{equation}
\hat{x}\approx  x_{1}sint+x_{2}cos t.
\end{equation}
Substituting equation (17) into equation (15) yields:
\begin{subequations}
	\begin{equation}
	x_{1}(1-r^2)-r \frac{1}{Q}x_2+\frac{3} {4}\eta_3 x_1 A^2 =0, 
	\end{equation}
	\begin{equation}
	x_{2}(1-r^2)+r \frac{1}{Q}x_1+\frac{3} {4}\eta_3 x_2 A^2 =\lambda,
	\end{equation}
\end{subequations}
where $A=\sqrt{x_{1}^2+x_{2}^2}$ is the amplitude of motion. Moreover, $x_{1}=A sin\phi$ and $x_{2}=A cos\phi$, $\phi$ being the phase difference between the excitation and the response. From equations (18a) and (18b) the following analytic frequency-amplitude relation could be found:
\begin{equation}
A^2 \Big[\Big((1-r^2)+\frac{3}{4}\eta_3 A^2\Big)^2+(\frac{r}{Q})^2\Big]=\lambda^2.
\end{equation}
\par
The idea of harmonic balance based parameter estimation is to follow a reverse path~\cite{amabili16}. In other words, the identification is conducted by assuming that the vibration amplitude $A$ and frequency ratio $r$  are already known for every frequency step from experiments. Therefore, in order to obtain unknown parameters, the following system should be solved for every $j$th frequency step, $r^{(j)}$:
\begin {equation}
\begin{pmatrix}
	-r^{(j)}x_2&\frac{3}{4}x_1 A^2 \\
	r^{(j)}x_1&\frac{3}{4}x_2 A^2
\end{pmatrix}
\cdot
\begin{bmatrix}
	\frac{1}{Q} \\
	\eta_3\\
\end{bmatrix}
=
\begin{bmatrix}
	-x_1(1-(r^{(j)})^2) \\
	-x_2(1-(r^{(j)})^2)+\lambda
\end{bmatrix}
\quad  ,   j=[1:m]
\end{equation}

System (20) can be compactly written as $\bar{A_{h}} \cdot X=\bar{B_{h}}$ . This system is over constrained since it contains  $2\times m$ equations. In order to solve system (20) and to estimate system parameters, least squares technique is used and the norm of the error $Er=\Big(\bar{A_{h}} \cdot X-\bar{B_{h}}\Big)\cdot\Big(\bar{A_{h}} \cdot X-\bar{B_{h}}\Big)^T $  should be minimized. Accordingly, here the pseudo-inverse of matrix $A_h$  is calculated and the solution is obtained as follows:
\begin{equation}
X=\Big(\bar{A_h}^T \bar{A_h}\Big)^{-1} \bar{A_h}\cdot \bar{B_h}.
\end{equation}
A problem in utilizing the least squares technique is that the identified peak amplitudes in the frequency-response curves  do not correspond to the ones obtained from the experiments. In order to resolve this issue, a correction on the quality factor is made by making use of the following expression (see~\cite{amabili16} for the derivation details):

\begin{equation}
Q=\frac{1}{2}\bigg[\sqrt{\frac{1}{2}+\frac{3}{8}\eta_3 A_{max}^2-\sqrt{(\frac{1}{2}+\frac{3}{8}\eta_3 A_{max}^2)^2-\frac{\lambda^2}{4 A_{max}^2}}}\bigg]^{-1}  ,
\end{equation}
in which $A_{max}$ is the experimentally measured peak amplitude for each frequency-amplitude curve. This will yield the nonlinear identification procedure to a single-fit parameter estimation algorithm for the estimation of $\eta_3$. 

\subsection*{3. Estimation of the electrostatic spring softening}

The electrostatic force acting on the membrane is given by

\begin{equation}
F_{el} = \frac{dU_{el}}{dx} = -\frac{1}{2}\frac{dC_g}{dx}V^2,
\end{equation}

\noindent where $U_{el} = -\frac{1}{2}C_g V^2$ is the electrostatic energy, $V = V_{dc} + V_{ac} cos(\omega t)$ is the applied voltage and $C_g$ is the gate capacitance. Assuming  $x<<g$, where $g$ is the gap between the membrane and the backgate, the gate capacitance can be approximated using a parallel plate capacitor model:

\begin{equation}
C_g = \varepsilon_0\frac{R^2 \pi}{g-x},
\end{equation}

\noindent where $R$ is the radius of the membrane and $\varepsilon_0$ is the vacuum permittivity. The resulting electrostatic force is given by

\begin{equation}
F = \frac{1}{2}\frac{\varepsilon_0 R^2\pi}{(g-x)^2}(V_{dc}+V_{ac}cos(\omega t))^2.
\end{equation}

If we expand this expression around $x = 0$, we get

\begin{equation}
F\approx \frac{1}{2}\varepsilon_0 R^2\pi(V_{dc}+V_{ac}cos(\omega t))^2 \big[\frac{1}{g^2}+\frac{2x}{g^3}+\frac{3x^2}{g^4}+\frac{4x^3}{g^5}\big].
\end{equation}

The first term ($\frac{1}{g^2}$) is the electrostatic actuation term and the second term is what is usually described as a spring softening term ($\frac{2x}{g^3}$). This term influences only the linear spring constant of the resonator. The term including $x^3$ will have a softening effect on the cubic spring constant: $k_{3,soft} = \frac{1}{2}\varepsilon_0 R^2\pi(V_{dc}+V_{ac}cos(\omega t))^2\frac{4x^3}{g^5} $. The resulting cubic spring constant $k_{3,tot}$ will be given by

\begin{equation}
k_{3,tot} = k_3 - k_{3,soft}.
\end{equation}

The ratio of the two contributions is

\begin{equation}
\frac{k_3}{k_{3,soft}} = \frac{1}{1.269-0.967 \nu - 0.269 \nu^2}\frac{Ehg^5}{2\varepsilon_0 R^4 (V_{dc}+V_{ac}cos(\omega t))^2}.
\end{equation}

For a Young's modulus $E = 594 GPa$, a radius of $R = 2.5 \mu m$, thickness of $h = 5 nm$ and a $V_{dc} < 3 V$, provided that $V_{dc}>>V_{ac}$, this ratio becomes

\begin{equation}
\label{ratio}
\frac{k_3}{k_{3,soft}} \approx 3000,
\end{equation}

\noindent which means that the electrostatic softening will have a negligible effect on the extracted Young's modulus (resulting in an error of $<0.1 \%$). It should be noted that the cavity depth has a significant influence on the effect of electrostatic softening of the cubic spring constant. To get resonable error margins ($< 5\%$), the ratio of eq.~(\ref{ratio}) should be kept above 20.

\subsection*{4. Nonlinear dynamic response as a function of Young's modulus}

In Fig. S2 we show the frequency response of the strongly-driven graphene drum (black dots) under a constant force ($F_\mathrm{RMS} = 48$ pN). The colored curves are the modeled response under constant force and with a fixed quality factor ($Q = $ 129) and resonance frequency $f_0 = 14.7$ MHz). The different colors correspond to the frequency responses of the model using different values for the Young's modulus to show how the nonlinear response is influenced by the Young's modulus.

\begin{figure}[ht]
	\includegraphics[width=0.45\textwidth]{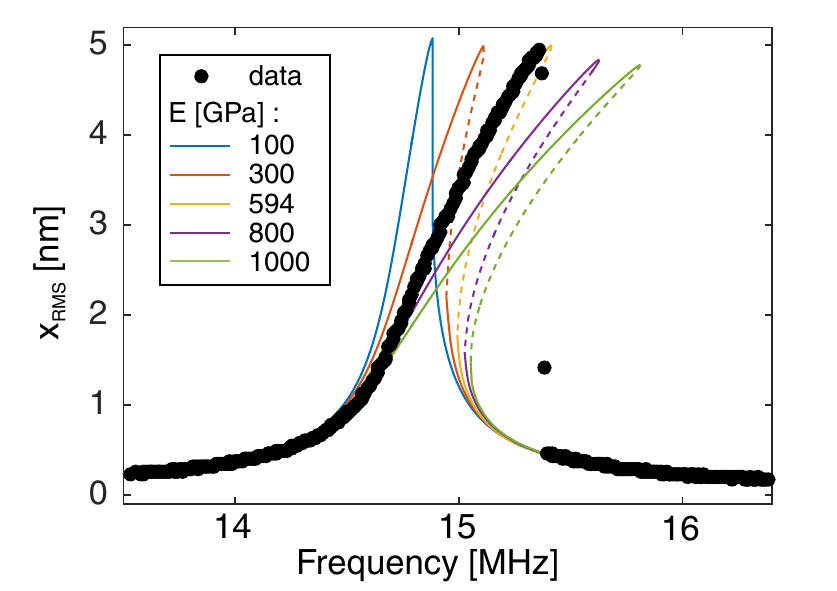}
	\caption{\label{fig:Fig3} Sensitivity of the nonlinear response to the Young's modulus. The measured trace (at $F_\mathrm{RMS} = 48$ pN) are represented by the black dots. The colored lines represent the modeled response using fixed values for the damping and the driving force and varying values for the Young's modulus. The yellow line represents the modeled response using the identified value for the Young's modulus (E = 594 GPa).}
\end{figure}


\begin{thebibliography}{39}%
	\makeatletter
	\providecommand \@ifxundefined [1]{%
		\@ifx{#1\undefined}
	}%
	\providecommand \@ifnum [1]{%
		\ifnum #1\expandafter \@firstoftwo
		\else \expandafter \@secondoftwo
		\fi
	}%
	\providecommand \@ifx [1]{%
		\ifx #1\expandafter \@firstoftwo
		\else \expandafter \@secondoftwo
		\fi
	}%
	\providecommand \natexlab [1]{#1}%
	\providecommand \enquote  [1]{``#1''}%
	\providecommand \bibnamefont  [1]{#1}%
	\providecommand \bibfnamefont [1]{#1}%
	\providecommand \citenamefont [1]{#1}%
	\providecommand \href@noop [0]{\@secondoftwo}%
	\providecommand \href [0]{\begingroup \@sanitize@url \@href}%
	\providecommand \@href[1]{\@@startlink{#1}\@@href}%
	\providecommand \@@href[1]{\endgroup#1\@@endlink}%
	\providecommand \@sanitize@url [0]{\catcode `\\12\catcode `\$12\catcode
		`\&12\catcode `\#12\catcode `\^12\catcode `\_12\catcode `\%12\relax}%
	\providecommand \@@startlink[1]{}%
	\providecommand \@@endlink[0]{}%
	\providecommand \url  [0]{\begingroup\@sanitize@url \@url }%
	\providecommand \@url [1]{\endgroup\@href {#1}{\urlprefix }}%
	\providecommand \urlprefix  [0]{URL }%
	\providecommand \Eprint [0]{\href }%
	\providecommand \doibase [0]{http://dx.doi.org/}%
	\providecommand \selectlanguage [0]{\@gobble}%
	\providecommand \bibinfo  [0]{\@secondoftwo}%
	\providecommand \bibfield  [0]{\@secondoftwo}%
	\providecommand \translation [1]{[#1]}%
	\providecommand \BibitemOpen [0]{}%
	\providecommand \bibitemStop [0]{}%
	\providecommand \bibitemNoStop [0]{.\EOS\space}%
	\providecommand \EOS [0]{\spacefactor3000\relax}%
	\providecommand \BibitemShut  [1]{\csname bibitem#1\endcsname}%
	\let\auto@bib@innerbib\@empty
	%</preamble>
	\bibitem [{\citenamefont {Smith}\ \emph {et~al.}(2013)\citenamefont {Smith},
		\citenamefont {Vaziri}, \citenamefont {Niklaus}, \citenamefont {Fischer},
		\citenamefont {Sterner}, \citenamefont {Delin}, \citenamefont {{\"O}stling},\
		and\ \citenamefont {Lemme}}]{smith13pressure}%
	\BibitemOpen
	\bibfield  {author} {\bibinfo {author} {\bibfnamefont {A.}~\bibnamefont
			{Smith}}, \bibinfo {author} {\bibfnamefont {S.}~\bibnamefont {Vaziri}},
		\bibinfo {author} {\bibfnamefont {F.}~\bibnamefont {Niklaus}}, \bibinfo
		{author} {\bibfnamefont {A.}~\bibnamefont {Fischer}}, \bibinfo {author}
		{\bibfnamefont {M.}~\bibnamefont {Sterner}}, \bibinfo {author} {\bibfnamefont
			{A.}~\bibnamefont {Delin}}, \bibinfo {author} {\bibfnamefont
			{M.}~\bibnamefont {{\"O}stling}}, \ and\ \bibinfo {author} {\bibfnamefont
			{M.}~\bibnamefont {Lemme}},\ }\href {\doibase
		http://dx.doi.org/10.1016/j.sse.2013.04.019} {\bibfield  {journal} {\bibinfo
			{journal} {Solid-State Electronics}\ }\textbf {\bibinfo {volume} {88}},\
		\bibinfo {pages} {89 } (\bibinfo {year} {2013})}\BibitemShut {NoStop}%
	\bibitem [{\citenamefont {Dolleman}\ \emph
		{et~al.}(2016{\natexlab{a}})\citenamefont {Dolleman}, \citenamefont
		{Davidovikj}, \citenamefont {Cartamil-Bueno}, \citenamefont {van~der Zant},\
		and\ \citenamefont {Steeneken}}]{dolleman15}%
	\BibitemOpen
	\bibfield  {author} {\bibinfo {author} {\bibfnamefont {R.~J.}\ \bibnamefont
			{Dolleman}}, \bibinfo {author} {\bibfnamefont {D.}~\bibnamefont
			{Davidovikj}}, \bibinfo {author} {\bibfnamefont {S.~J.}\ \bibnamefont
			{Cartamil-Bueno}}, \bibinfo {author} {\bibfnamefont {H.~S.~J.}\ \bibnamefont
			{van~der Zant}}, \ and\ \bibinfo {author} {\bibfnamefont {P.~G.}\
			\bibnamefont {Steeneken}},\ }\href@noop {} {\bibfield  {journal} {\bibinfo
			{journal} {Nano Letters}\ }\textbf {\bibinfo {volume} {16}},\ \bibinfo
		{pages} {568} (\bibinfo {year} {2016}{\natexlab{a}})}\BibitemShut {NoStop}%
	\bibitem [{\citenamefont {Koenig}\ \emph {et~al.}(2012)\citenamefont {Koenig},
		\citenamefont {Wang}, \citenamefont {Pellegrino},\ and\ \citenamefont
		{Bunch}}]{bunch12}%
	\BibitemOpen
	\bibfield  {author} {\bibinfo {author} {\bibfnamefont {S.~P.}\ \bibnamefont
			{Koenig}}, \bibinfo {author} {\bibfnamefont {L.}~\bibnamefont {Wang}},
		\bibinfo {author} {\bibfnamefont {J.}~\bibnamefont {Pellegrino}}, \ and\
		\bibinfo {author} {\bibfnamefont {J.~S.}\ \bibnamefont {Bunch}},\ }\href@noop
	{} {\bibfield  {journal} {\bibinfo  {journal} {Nature Nanotechnology}\
		}\textbf {\bibinfo {volume} {7}},\ \bibinfo {pages} {728} (\bibinfo {year}
		{2012})}\BibitemShut {NoStop}%
	\bibitem [{\citenamefont {Dolleman}\ \emph
		{et~al.}(2016{\natexlab{b}})\citenamefont {Dolleman}, \citenamefont
		{Cartamil-Bueno}, \citenamefont {van~der Zant},\ and\ \citenamefont
		{Steeneken}}]{dolleman16}%
	\BibitemOpen
	\bibfield  {author} {\bibinfo {author} {\bibfnamefont {R.~J.}\ \bibnamefont
			{Dolleman}}, \bibinfo {author} {\bibfnamefont {S.~J.}\ \bibnamefont
			{Cartamil-Bueno}}, \bibinfo {author} {\bibfnamefont {H.~S.~J.}\ \bibnamefont
			{van~der Zant}}, \ and\ \bibinfo {author} {\bibfnamefont {P.~G.}\
			\bibnamefont {Steeneken}},\ }\href
	{http://stacks.iop.org/2053-1583/4/i=1/a=011002} {\bibfield  {journal}
		{\bibinfo  {journal} {2D Materials}\ }\textbf {\bibinfo {volume} {4}},\
		\bibinfo {pages} {011002} (\bibinfo {year} {2016}{\natexlab{b}})}\BibitemShut
	{NoStop}%
	\bibitem [{\citenamefont {Sakhaee-Pour}\ \emph {et~al.}(2008)\citenamefont
		{Sakhaee-Pour}, \citenamefont {Ahmadian},\ and\ \citenamefont
		{Vafai}}]{sakhaee08mass}%
	\BibitemOpen
	\bibfield  {author} {\bibinfo {author} {\bibfnamefont {A.}~\bibnamefont
			{Sakhaee-Pour}}, \bibinfo {author} {\bibfnamefont {M.}~\bibnamefont
			{Ahmadian}}, \ and\ \bibinfo {author} {\bibfnamefont {A.}~\bibnamefont
			{Vafai}},\ }\href@noop {} {\bibfield  {journal} {\bibinfo  {journal} {Solid
				State Communications}\ }\textbf {\bibinfo {volume} {145}},\ \bibinfo {pages}
		{168} (\bibinfo {year} {2008})}\BibitemShut {NoStop}%
	\bibitem [{\citenamefont {Atalaya}\ \emph {et~al.}(2010)\citenamefont
		{Atalaya}, \citenamefont {Kinaret},\ and\ \citenamefont
		{Isacsson}}]{atalaya10}%
	\BibitemOpen
	\bibfield  {author} {\bibinfo {author} {\bibfnamefont {J.}~\bibnamefont
			{Atalaya}}, \bibinfo {author} {\bibfnamefont {J.~M.}\ \bibnamefont
			{Kinaret}}, \ and\ \bibinfo {author} {\bibfnamefont {A.}~\bibnamefont
			{Isacsson}},\ }\href@noop {} {\bibfield  {journal} {\bibinfo  {journal} {EPL
				(Europhysics Letters)}\ }\textbf {\bibinfo {volume} {91}},\ \bibinfo {pages}
		{48001} (\bibinfo {year} {2010})}\BibitemShut {NoStop}%
	\bibitem [{\citenamefont {Lee}\ \emph {et~al.}(2008)\citenamefont {Lee},
		\citenamefont {Wei}, \citenamefont {Kysar},\ and\ \citenamefont
		{Hone}}]{hone08elastic}%
	\BibitemOpen
	\bibfield  {author} {\bibinfo {author} {\bibfnamefont {C.}~\bibnamefont
			{Lee}}, \bibinfo {author} {\bibfnamefont {X.}~\bibnamefont {Wei}}, \bibinfo
		{author} {\bibfnamefont {J.~W.}\ \bibnamefont {Kysar}}, \ and\ \bibinfo
		{author} {\bibfnamefont {J.}~\bibnamefont {Hone}},\ }\href {\doibase
		10.1126/science.1157996} {\bibfield  {journal} {\bibinfo  {journal}
			{Science}\ }\textbf {\bibinfo {volume} {321}},\ \bibinfo {pages} {385}
		(\bibinfo {year} {2008})}\BibitemShut {NoStop}%
	\bibitem [{\citenamefont {Poot}\ and\ \citenamefont {van~der
			Zant}(2008)}]{poot08}%
	\BibitemOpen
	\bibfield  {author} {\bibinfo {author} {\bibfnamefont {M.}~\bibnamefont
			{Poot}}\ and\ \bibinfo {author} {\bibfnamefont {H.~S.~J.}\ \bibnamefont
			{van~der Zant}},\ }\href@noop {} {\bibfield  {journal} {\bibinfo  {journal}
			{Applied Physics Letters}\ }\textbf {\bibinfo {volume} {92}},\ \bibinfo
		{pages} {063111} (\bibinfo {year} {2008})}\BibitemShut {NoStop}%
	\bibitem [{\citenamefont {Castellanos-Gomez}\ \emph
		{et~al.}(2012{\natexlab{a}})\citenamefont {Castellanos-Gomez}, \citenamefont
		{Poot}, \citenamefont {Steele}, \citenamefont {van~der Zant}, \citenamefont
		{Agra{\"\i}t},\ and\ \citenamefont {Rubio-Bollinger}}]{castellanos12}%
	\BibitemOpen
	\bibfield  {author} {\bibinfo {author} {\bibfnamefont {A.}~\bibnamefont
			{Castellanos-Gomez}}, \bibinfo {author} {\bibfnamefont {M.}~\bibnamefont
			{Poot}}, \bibinfo {author} {\bibfnamefont {G.~A.}\ \bibnamefont {Steele}},
		\bibinfo {author} {\bibfnamefont {H.~S.~J.}\ \bibnamefont {van~der Zant}},
		\bibinfo {author} {\bibfnamefont {N.}~\bibnamefont {Agra{\"\i}t}}, \ and\
		\bibinfo {author} {\bibfnamefont {G.}~\bibnamefont {Rubio-Bollinger}},\
	}\href@noop {} {\bibfield  {journal} {\bibinfo  {journal} {Nanoscale Research
			Letters}\ }\textbf {\bibinfo {volume} {7}},\ \bibinfo {pages} {1} (\bibinfo
	{year} {2012}{\natexlab{a}})}\BibitemShut {NoStop}%
\bibitem [{\citenamefont {Castellanos-Gomez}\ \emph
	{et~al.}(2012{\natexlab{b}})\citenamefont {Castellanos-Gomez}, \citenamefont
	{Poot}, \citenamefont {Steele}, \citenamefont {van~der Zant}, \citenamefont
	{Agra{\"\i}t},\ and\ \citenamefont {Rubio-Bollinger}}]{castellanos12am}%
\BibitemOpen
\bibfield  {author} {\bibinfo {author} {\bibfnamefont {A.}~\bibnamefont
		{Castellanos-Gomez}}, \bibinfo {author} {\bibfnamefont {M.}~\bibnamefont
		{Poot}}, \bibinfo {author} {\bibfnamefont {G.~A.}\ \bibnamefont {Steele}},
	\bibinfo {author} {\bibfnamefont {H.~S.~J.}\ \bibnamefont {van~der Zant}},
	\bibinfo {author} {\bibfnamefont {N.}~\bibnamefont {Agra{\"\i}t}}, \ and\
	\bibinfo {author} {\bibfnamefont {G.}~\bibnamefont {Rubio-Bollinger}},\
}\href@noop {} {\bibfield  {journal} {\bibinfo  {journal} {Advanced
		Materials}\ }\textbf {\bibinfo {volume} {24}},\ \bibinfo {pages} {772}
(\bibinfo {year} {2012}{\natexlab{b}})}\BibitemShut {NoStop}%
\bibitem [{\citenamefont {Koenig}\ \emph {et~al.}(2011)\citenamefont {Koenig},
	\citenamefont {Boddeti}, \citenamefont {Dunn},\ and\ \citenamefont
	{Bunch}}]{koenig11}%
\BibitemOpen
\bibfield  {author} {\bibinfo {author} {\bibfnamefont {S.~P.}\ \bibnamefont
		{Koenig}}, \bibinfo {author} {\bibfnamefont {N.~G.}\ \bibnamefont {Boddeti}},
	\bibinfo {author} {\bibfnamefont {M.~L.}\ \bibnamefont {Dunn}}, \ and\
	\bibinfo {author} {\bibfnamefont {J.~S.}\ \bibnamefont {Bunch}},\ }\href@noop
{} {\bibfield  {journal} {\bibinfo  {journal} {Nature nanotechnology}\
	}\textbf {\bibinfo {volume} {6}},\ \bibinfo {pages} {543} (\bibinfo {year}
	{2011})}\BibitemShut {NoStop}%
\bibitem [{\citenamefont {Wong}\ \emph {et~al.}(2010)\citenamefont {Wong},
	\citenamefont {Annamalai}, \citenamefont {Wang},\ and\ \citenamefont
	{Palaniapan}}]{wong10}%
\BibitemOpen
\bibfield  {author} {\bibinfo {author} {\bibfnamefont {C.}~\bibnamefont
		{Wong}}, \bibinfo {author} {\bibfnamefont {M.}~\bibnamefont {Annamalai}},
	\bibinfo {author} {\bibfnamefont {Z.}~\bibnamefont {Wang}}, \ and\ \bibinfo
	{author} {\bibfnamefont {M.}~\bibnamefont {Palaniapan}},\ }\href@noop {}
{\bibfield  {journal} {\bibinfo  {journal} {Journal of Micromechanics and
			Microengineering}\ }\textbf {\bibinfo {volume} {20}},\ \bibinfo {pages}
	{115029} (\bibinfo {year} {2010})}\BibitemShut {NoStop}%
\bibitem [{\citenamefont {Nicholl}\ \emph {et~al.}(2015)\citenamefont
	{Nicholl}, \citenamefont {Conley}, \citenamefont {Lavrik}, \citenamefont
	{Vlassiouk}, \citenamefont {Puzyrev}, \citenamefont {Sreenivas},
	\citenamefont {Pantelides},\ and\ \citenamefont {Bolotin}}]{nicholl15}%
\BibitemOpen
\bibfield  {author} {\bibinfo {author} {\bibfnamefont {R.~J.}\ \bibnamefont
		{Nicholl}}, \bibinfo {author} {\bibfnamefont {H.~J.}\ \bibnamefont {Conley}},
	\bibinfo {author} {\bibfnamefont {N.~V.}\ \bibnamefont {Lavrik}}, \bibinfo
	{author} {\bibfnamefont {I.}~\bibnamefont {Vlassiouk}}, \bibinfo {author}
	{\bibfnamefont {Y.~S.}\ \bibnamefont {Puzyrev}}, \bibinfo {author}
	{\bibfnamefont {V.~P.}\ \bibnamefont {Sreenivas}}, \bibinfo {author}
	{\bibfnamefont {S.~T.}\ \bibnamefont {Pantelides}}, \ and\ \bibinfo {author}
	{\bibfnamefont {K.~I.}\ \bibnamefont {Bolotin}},\ }\href@noop {} {\bibfield
	{journal} {\bibinfo  {journal} {Nature Communications}\ }\textbf {\bibinfo
		{volume} {6}} (\bibinfo {year} {2015})}\BibitemShut {NoStop}%
\bibitem [{\citenamefont {Bunch}\ \emph {et~al.}(2007)\citenamefont {Bunch},
	\citenamefont {Van Der~Zande}, \citenamefont {Verbridge}, \citenamefont
	{Frank}, \citenamefont {Tanenbaum}, \citenamefont {Parpia}, \citenamefont
	{Craighead},\ and\ \citenamefont {McEuen}}]{bunch07}%
\BibitemOpen
\bibfield  {author} {\bibinfo {author} {\bibfnamefont {J.~S.}\ \bibnamefont
		{Bunch}}, \bibinfo {author} {\bibfnamefont {A.~M.}\ \bibnamefont {Van
			Der~Zande}}, \bibinfo {author} {\bibfnamefont {S.~S.}\ \bibnamefont
		{Verbridge}}, \bibinfo {author} {\bibfnamefont {I.~W.}\ \bibnamefont
		{Frank}}, \bibinfo {author} {\bibfnamefont {D.~M.}\ \bibnamefont
		{Tanenbaum}}, \bibinfo {author} {\bibfnamefont {J.~M.}\ \bibnamefont
		{Parpia}}, \bibinfo {author} {\bibfnamefont {H.~G.}\ \bibnamefont
		{Craighead}}, \ and\ \bibinfo {author} {\bibfnamefont {P.~L.}\ \bibnamefont
		{McEuen}},\ }\href@noop {} {\bibfield  {journal} {\bibinfo  {journal}
		{Science}\ }\textbf {\bibinfo {volume} {315}},\ \bibinfo {pages} {490}
	(\bibinfo {year} {2007})}\BibitemShut {NoStop}%
\bibitem [{\citenamefont {Castellanos-Gomez}\ \emph {et~al.}(2013)\citenamefont
	{Castellanos-Gomez}, \citenamefont {van Leeuwen}, \citenamefont {Buscema},
	\citenamefont {van~der Zant}, \citenamefont {Steele},\ and\ \citenamefont
	{Venstra}}]{castellanos13}%
\BibitemOpen
\bibfield  {author} {\bibinfo {author} {\bibfnamefont {A.}~\bibnamefont
		{Castellanos-Gomez}}, \bibinfo {author} {\bibfnamefont {R.}~\bibnamefont {van
			Leeuwen}}, \bibinfo {author} {\bibfnamefont {M.}~\bibnamefont {Buscema}},
	\bibinfo {author} {\bibfnamefont {H.~S.~J.}\ \bibnamefont {van~der Zant}},
	\bibinfo {author} {\bibfnamefont {G.~A.}\ \bibnamefont {Steele}}, \ and\
	\bibinfo {author} {\bibfnamefont {W.~J.}\ \bibnamefont {Venstra}},\
}\href@noop {} {\bibfield  {journal} {\bibinfo  {journal} {Advanced
		Materials}\ }\textbf {\bibinfo {volume} {25}},\ \bibinfo {pages} {6719}
(\bibinfo {year} {2013})}\BibitemShut {NoStop}%
\bibitem [{\citenamefont {Davidovikj}\ \emph {et~al.}(2016)\citenamefont
	{Davidovikj}, \citenamefont {Slim}, \citenamefont {Cartamil-Bueno},
	\citenamefont {van~der Zant}, \citenamefont {Steeneken},\ and\ \citenamefont
	{Venstra}}]{davidovikj16}%
\BibitemOpen
\bibfield  {author} {\bibinfo {author} {\bibfnamefont {D.}~\bibnamefont
		{Davidovikj}}, \bibinfo {author} {\bibfnamefont {J.~J.}\ \bibnamefont
		{Slim}}, \bibinfo {author} {\bibfnamefont {S.~J.}\ \bibnamefont
		{Cartamil-Bueno}}, \bibinfo {author} {\bibfnamefont {H.~S.}\ \bibnamefont
		{van~der Zant}}, \bibinfo {author} {\bibfnamefont {P.~G.}\ \bibnamefont
		{Steeneken}}, \ and\ \bibinfo {author} {\bibfnamefont {W.~J.}\ \bibnamefont
		{Venstra}},\ }\href@noop {} {\bibfield  {journal} {\bibinfo  {journal} {Nano
			letters}\ }\textbf {\bibinfo {volume} {16}},\ \bibinfo {pages} {2768}
	(\bibinfo {year} {2016})}\BibitemShut {NoStop}%
\bibitem [{\citenamefont {Cartamil-Bueno}\ \emph {et~al.}(2015)\citenamefont
	{Cartamil-Bueno}, \citenamefont {Steeneken}, \citenamefont {Tichelaar},
	\citenamefont {Navarro-Moratalla}, \citenamefont {Venstra}, \citenamefont
	{van Leeuwen}, \citenamefont {Coronado}, \citenamefont {van~der Zant},
	\citenamefont {Steele},\ and\ \citenamefont
	{Castellanos-Gomez}}]{cartamil15}%
\BibitemOpen
\bibfield  {author} {\bibinfo {author} {\bibfnamefont {S.~J.}\ \bibnamefont
		{Cartamil-Bueno}}, \bibinfo {author} {\bibfnamefont {P.~G.}\ \bibnamefont
		{Steeneken}}, \bibinfo {author} {\bibfnamefont {F.~D.}\ \bibnamefont
		{Tichelaar}}, \bibinfo {author} {\bibfnamefont {E.}~\bibnamefont
		{Navarro-Moratalla}}, \bibinfo {author} {\bibfnamefont {W.~J.}\ \bibnamefont
		{Venstra}}, \bibinfo {author} {\bibfnamefont {R.}~\bibnamefont {van
			Leeuwen}}, \bibinfo {author} {\bibfnamefont {E.}~\bibnamefont {Coronado}},
	\bibinfo {author} {\bibfnamefont {H.~S.~J.}\ \bibnamefont {van~der Zant}},
	\bibinfo {author} {\bibfnamefont {G.~A.}\ \bibnamefont {Steele}}, \ and\
	\bibinfo {author} {\bibfnamefont {A.}~\bibnamefont {Castellanos-Gomez}},\
}\href@noop {} {\bibfield  {journal} {\bibinfo  {journal} {Nano Research}\
}\textbf {\bibinfo {volume} {8}} (\bibinfo {year} {2015})}\BibitemShut
{NoStop}%
\bibitem [{\citenamefont {Wang}\ \emph {et~al.}(2015)\citenamefont {Wang},
	\citenamefont {Jia}, \citenamefont {Zheng}, \citenamefont {Yang},
	\citenamefont {Wang}, \citenamefont {Ye}, \citenamefont {Chen}, \citenamefont
	{Shan},\ and\ \citenamefont {Feng}}]{wang15}%
\BibitemOpen
\bibfield  {author} {\bibinfo {author} {\bibfnamefont {Z.}~\bibnamefont
		{Wang}}, \bibinfo {author} {\bibfnamefont {H.}~\bibnamefont {Jia}}, \bibinfo
	{author} {\bibfnamefont {X.}~\bibnamefont {Zheng}}, \bibinfo {author}
	{\bibfnamefont {R.}~\bibnamefont {Yang}}, \bibinfo {author} {\bibfnamefont
		{Z.}~\bibnamefont {Wang}}, \bibinfo {author} {\bibfnamefont {G.}~\bibnamefont
		{Ye}}, \bibinfo {author} {\bibfnamefont {X.}~\bibnamefont {Chen}}, \bibinfo
	{author} {\bibfnamefont {J.}~\bibnamefont {Shan}}, \ and\ \bibinfo {author}
	{\bibfnamefont {P.~X.-L.}\ \bibnamefont {Feng}},\ }\href@noop {} {\bibfield
	{journal} {\bibinfo  {journal} {Nanoscale}\ }\textbf {\bibinfo {volume}
		{7}},\ \bibinfo {pages} {877} (\bibinfo {year} {2015})}\BibitemShut {NoStop}%
\bibitem [{\citenamefont {Eichler}\ \emph {et~al.}(2011)\citenamefont
	{Eichler}, \citenamefont {Moser}, \citenamefont {Chaste}, \citenamefont
	{Zdrojek}, \citenamefont {Wilson-Rae},\ and\ \citenamefont
	{Bachtold}}]{eichler2011}%
\BibitemOpen
\bibfield  {author} {\bibinfo {author} {\bibfnamefont {A.}~\bibnamefont
		{Eichler}}, \bibinfo {author} {\bibfnamefont {J.}~\bibnamefont {Moser}},
	\bibinfo {author} {\bibfnamefont {J.}~\bibnamefont {Chaste}}, \bibinfo
	{author} {\bibfnamefont {M.}~\bibnamefont {Zdrojek}}, \bibinfo {author}
	{\bibfnamefont {I.}~\bibnamefont {Wilson-Rae}}, \ and\ \bibinfo {author}
	{\bibfnamefont {A.}~\bibnamefont {Bachtold}},\ }\href@noop {} {\bibfield
	{journal} {\bibinfo  {journal} {Nature nanotechnology}\ }\textbf {\bibinfo
		{volume} {6}},\ \bibinfo {pages} {339} (\bibinfo {year} {2011})}\BibitemShut
{NoStop}%
\bibitem [{\citenamefont {Croy}\ \emph {et~al.}(2012)\citenamefont {Croy},
	\citenamefont {Midtvedt}, \citenamefont {Isacsson},\ and\ \citenamefont
	{Kinaret}}]{croy12}%
\BibitemOpen
\bibfield  {author} {\bibinfo {author} {\bibfnamefont {A.}~\bibnamefont
		{Croy}}, \bibinfo {author} {\bibfnamefont {D.}~\bibnamefont {Midtvedt}},
	\bibinfo {author} {\bibfnamefont {A.}~\bibnamefont {Isacsson}}, \ and\
	\bibinfo {author} {\bibfnamefont {J.~M.}\ \bibnamefont {Kinaret}},\ }\href
{\doibase 10.1103/PhysRevB.86.235435} {\bibfield  {journal} {\bibinfo
		{journal} {Phys. Rev. B}\ }\textbf {\bibinfo {volume} {86}},\ \bibinfo
	{pages} {235435} (\bibinfo {year} {2012})}\BibitemShut {NoStop}%
\bibitem [{\citenamefont {Eriksson}\ \emph {et~al.}(2013)\citenamefont
	{Eriksson}, \citenamefont {Midtvedt}, \citenamefont {Croy},\ and\
	\citenamefont {Isacsson}}]{eriksson13}%
\BibitemOpen
\bibfield  {author} {\bibinfo {author} {\bibfnamefont {A.}~\bibnamefont
		{Eriksson}}, \bibinfo {author} {\bibfnamefont {D.}~\bibnamefont {Midtvedt}},
	\bibinfo {author} {\bibfnamefont {A.}~\bibnamefont {Croy}}, \ and\ \bibinfo
	{author} {\bibfnamefont {A.}~\bibnamefont {Isacsson}},\ }\href@noop {}
{\bibfield  {journal} {\bibinfo  {journal} {Nanotechnology}\ }\textbf
	{\bibinfo {volume} {24}},\ \bibinfo {pages} {395702} (\bibinfo {year}
	{2013})}\BibitemShut {NoStop}%
\bibitem [{\citenamefont {De~Alba}\ \emph {et~al.}(2016)\citenamefont
	{De~Alba}, \citenamefont {Massel}, \citenamefont {Storch}, \citenamefont
	{Abhilash}, \citenamefont {Hui}, \citenamefont {McEuen}, \citenamefont
	{Craighead},\ and\ \citenamefont {Parpia}}]{dealba16}%
\BibitemOpen
\bibfield  {author} {\bibinfo {author} {\bibfnamefont {R.}~\bibnamefont
		{De~Alba}}, \bibinfo {author} {\bibfnamefont {F.}~\bibnamefont {Massel}},
	\bibinfo {author} {\bibfnamefont {I.}~\bibnamefont {Storch}}, \bibinfo
	{author} {\bibfnamefont {T.}~\bibnamefont {Abhilash}}, \bibinfo {author}
	{\bibfnamefont {A.}~\bibnamefont {Hui}}, \bibinfo {author} {\bibfnamefont
		{P.}~\bibnamefont {McEuen}}, \bibinfo {author} {\bibfnamefont
		{H.}~\bibnamefont {Craighead}}, \ and\ \bibinfo {author} {\bibfnamefont
		{J.}~\bibnamefont {Parpia}},\ }\href@noop {} {\bibfield  {journal} {\bibinfo
		{journal} {Nature Nanotechnology}\ }\textbf {\bibinfo {volume} {11}},\
	\bibinfo {pages} {741} (\bibinfo {year} {2016})}\BibitemShut {NoStop}%
\bibitem [{\citenamefont {Mathew}\ \emph {et~al.}(2016)\citenamefont {Mathew},
	\citenamefont {Patel}, \citenamefont {Borah}, \citenamefont {Vijay},\ and\
	\citenamefont {Deshmukh}}]{mathew16}%
\BibitemOpen
\bibfield  {author} {\bibinfo {author} {\bibfnamefont {J.~P.}\ \bibnamefont
		{Mathew}}, \bibinfo {author} {\bibfnamefont {R.~N.}\ \bibnamefont {Patel}},
	\bibinfo {author} {\bibfnamefont {A.}~\bibnamefont {Borah}}, \bibinfo
	{author} {\bibfnamefont {R.}~\bibnamefont {Vijay}}, \ and\ \bibinfo {author}
	{\bibfnamefont {M.~M.}\ \bibnamefont {Deshmukh}},\ }\href@noop {} {\bibfield
	{journal} {\bibinfo  {journal} {Nature Nanotechnology}\ }\textbf {\bibinfo
		{volume} {11}},\ \bibinfo {pages} {747} (\bibinfo {year} {2016})}\BibitemShut
{NoStop}%
\bibitem [{\citenamefont {Chen}\ \emph {et~al.}(2009)\citenamefont {Chen},
	\citenamefont {Rosenblatt}, \citenamefont {Bolotin}, \citenamefont {Kalb},
	\citenamefont {Kim}, \citenamefont {Kymissis}, \citenamefont {Stormer},
	\citenamefont {Heinz},\ and\ \citenamefont {Hone}}]{chen09}%
\BibitemOpen
\bibfield  {author} {\bibinfo {author} {\bibfnamefont {C.}~\bibnamefont
		{Chen}}, \bibinfo {author} {\bibfnamefont {S.}~\bibnamefont {Rosenblatt}},
	\bibinfo {author} {\bibfnamefont {K.~I.}\ \bibnamefont {Bolotin}}, \bibinfo
	{author} {\bibfnamefont {W.}~\bibnamefont {Kalb}}, \bibinfo {author}
	{\bibfnamefont {P.}~\bibnamefont {Kim}}, \bibinfo {author} {\bibfnamefont
		{I.}~\bibnamefont {Kymissis}}, \bibinfo {author} {\bibfnamefont {H.~L.}\
		\bibnamefont {Stormer}}, \bibinfo {author} {\bibfnamefont {T.~F.}\
		\bibnamefont {Heinz}}, \ and\ \bibinfo {author} {\bibfnamefont
		{J.}~\bibnamefont {Hone}},\ }\href@noop {} {\bibfield  {journal} {\bibinfo
		{journal} {Nature Nanotechnology}\ }\textbf {\bibinfo {volume} {4}},\
	\bibinfo {pages} {861} (\bibinfo {year} {2009})}\BibitemShut {NoStop}%
\bibitem [{\citenamefont {Chen}\ \emph {et~al.}(2013)\citenamefont {Chen},
	\citenamefont {Lee}, \citenamefont {Deshpande}, \citenamefont {Lee},
	\citenamefont {Lekas}, \citenamefont {Shepard},\ and\ \citenamefont
	{Hone}}]{chen13}%
\BibitemOpen
\bibfield  {author} {\bibinfo {author} {\bibfnamefont {C.}~\bibnamefont
		{Chen}}, \bibinfo {author} {\bibfnamefont {S.}~\bibnamefont {Lee}}, \bibinfo
	{author} {\bibfnamefont {V.~V.}\ \bibnamefont {Deshpande}}, \bibinfo {author}
	{\bibfnamefont {G.-H.}\ \bibnamefont {Lee}}, \bibinfo {author} {\bibfnamefont
		{M.}~\bibnamefont {Lekas}}, \bibinfo {author} {\bibfnamefont
		{K.}~\bibnamefont {Shepard}}, \ and\ \bibinfo {author} {\bibfnamefont
		{J.}~\bibnamefont {Hone}},\ }\href@noop {} {\bibfield  {journal} {\bibinfo
		{journal} {Nature Nanotechnology}\ }\textbf {\bibinfo {volume} {8}},\
	\bibinfo {pages} {923} (\bibinfo {year} {2013})}\BibitemShut {NoStop}%
\bibitem [{\citenamefont {Han}\ \emph {et~al.}(2015)\citenamefont {Han},
	\citenamefont {Pugno},\ and\ \citenamefont {Ryu}}]{han15}%
\BibitemOpen
\bibfield  {author} {\bibinfo {author} {\bibfnamefont {J.}~\bibnamefont
		{Han}}, \bibinfo {author} {\bibfnamefont {N.~M.}\ \bibnamefont {Pugno}}, \
	and\ \bibinfo {author} {\bibfnamefont {S.}~\bibnamefont {Ryu}},\ }\href
{\doibase 10.1039/C5NR04134A} {\bibfield  {journal} {\bibinfo  {journal}
		{Nanoscale}\ }\textbf {\bibinfo {volume} {7}},\ \bibinfo {pages} {15672}
	(\bibinfo {year} {2015})}\BibitemShut {NoStop}%
\bibitem [{\citenamefont {Vella}\ and\ \citenamefont
	{Davidovitch}(2017)}]{vella2017}%
\BibitemOpen
\bibfield  {author} {\bibinfo {author} {\bibfnamefont {D.}~\bibnamefont
		{Vella}}\ and\ \bibinfo {author} {\bibfnamefont {B.}~\bibnamefont
		{Davidovitch}},\ }\href@noop {} {\bibfield  {journal} {\bibinfo  {journal}
		{Soft Matter}\ } (\bibinfo {year} {2017})}\BibitemShut {NoStop}%
\bibitem [{\citenamefont {Castellanos-Gomez}\ \emph {et~al.}(2014)\citenamefont
	{Castellanos-Gomez}, \citenamefont {Buscema}, \citenamefont {Molenaar},
	\citenamefont {Singh}, \citenamefont {Janssen}, \citenamefont {van~der
		Zant},\ and\ \citenamefont {Steele}}]{castellanos14}%
\BibitemOpen
\bibfield  {author} {\bibinfo {author} {\bibfnamefont {A.}~\bibnamefont
		{Castellanos-Gomez}}, \bibinfo {author} {\bibfnamefont {M.}~\bibnamefont
		{Buscema}}, \bibinfo {author} {\bibfnamefont {R.}~\bibnamefont {Molenaar}},
	\bibinfo {author} {\bibfnamefont {V.}~\bibnamefont {Singh}}, \bibinfo
	{author} {\bibfnamefont {L.}~\bibnamefont {Janssen}}, \bibinfo {author}
	{\bibfnamefont {H.~S.~J.}\ \bibnamefont {van~der Zant}}, \ and\ \bibinfo
	{author} {\bibfnamefont {G.~A.}\ \bibnamefont {Steele}},\ }\href@noop {}
{\bibfield  {journal} {\bibinfo  {journal} {2D Materials}\ }\textbf {\bibinfo
		{volume} {1}},\ \bibinfo {pages} {011002} (\bibinfo {year}
	{2014})}\BibitemShut {NoStop}%
\bibitem [{\citenamefont {Amabili}\ \emph {et~al.}(2016)\citenamefont
	{Amabili}, \citenamefont {Alijani},\ and\ \citenamefont
	{Delannoy}}]{amabili16}%
\BibitemOpen
\bibfield  {author} {\bibinfo {author} {\bibfnamefont {M.}~\bibnamefont
		{Amabili}}, \bibinfo {author} {\bibfnamefont {F.}~\bibnamefont {Alijani}}, \
	and\ \bibinfo {author} {\bibfnamefont {J.}~\bibnamefont {Delannoy}},\
}\href@noop {} {\bibfield  {journal} {\bibinfo  {journal} {International
		Journal of Non-Linear Mechanics}\ }\textbf {\bibinfo {volume} {85}},\
\bibinfo {pages} {226} (\bibinfo {year} {2016})}\BibitemShut {NoStop}%
\bibitem [{\citenamefont {Singh}\ \emph {et~al.}(2016)\citenamefont {Singh},
	\citenamefont {Shevchuk}, \citenamefont {Blanter},\ and\ \citenamefont
	{Steele}}]{singh16}%
\BibitemOpen
\bibfield  {author} {\bibinfo {author} {\bibfnamefont {V.}~\bibnamefont
		{Singh}}, \bibinfo {author} {\bibfnamefont {O.}~\bibnamefont {Shevchuk}},
	\bibinfo {author} {\bibfnamefont {Y.~M.}\ \bibnamefont {Blanter}}, \ and\
	\bibinfo {author} {\bibfnamefont {G.~A.}\ \bibnamefont {Steele}},\ }\href
{\doibase 10.1103/PhysRevB.93.245407} {\bibfield  {journal} {\bibinfo
		{journal} {Phys. Rev. B}\ }\textbf {\bibinfo {volume} {93}},\ \bibinfo
	{pages} {245407} (\bibinfo {year} {2016})}\BibitemShut {NoStop}%
\bibitem [{\citenamefont {Mansfield}(2005)}]{mansfield2005}%
\BibitemOpen
\bibfield  {author} {\bibinfo {author} {\bibfnamefont {E.~H.}\ \bibnamefont
		{Mansfield}},\ }\href@noop {} {\emph {\bibinfo {title} {The bending and
			stretching of plates}}}\ (\bibinfo  {publisher} {Cambridge University
	Press},\ \bibinfo {year} {2005})\BibitemShut {NoStop}%
\bibitem [{\citenamefont {Komaragiri}\ \emph {et~al.}(2005)\citenamefont
	{Komaragiri}, \citenamefont {Begley},\ and\ \citenamefont
	{Simmonds}}]{komaragiri05}%
\BibitemOpen
\bibfield  {author} {\bibinfo {author} {\bibfnamefont {U.}~\bibnamefont
		{Komaragiri}}, \bibinfo {author} {\bibfnamefont {M.}~\bibnamefont {Begley}},
	\ and\ \bibinfo {author} {\bibfnamefont {J.}~\bibnamefont {Simmonds}},\
}\href@noop {} {\bibfield  {journal} {\bibinfo  {journal} {Transactions of
		the ASME-E-Journal of Applied Mechanics}\ }\textbf {\bibinfo {volume} {72}},\
\bibinfo {pages} {203} (\bibinfo {year} {2005})}\BibitemShut {NoStop}%
\bibitem [{\citenamefont {Hencky}(1915)}]{hencky15}%
\BibitemOpen
\bibfield  {author} {\bibinfo {author} {\bibfnamefont {H.}~\bibnamefont
		{Hencky}},\ }\href@noop {} {\bibfield  {journal} {\bibinfo  {journal}
		{Zeitschrift fur Mathematik und Physik}\ }\textbf {\bibinfo {volume} {63}},\
	\bibinfo {pages} {311} (\bibinfo {year} {1915})}\BibitemShut {NoStop}%
\bibitem [{\citenamefont {Boddeti}\ \emph {et~al.}(2013)\citenamefont
	{Boddeti}, \citenamefont {Koenig}, \citenamefont {Long}, \citenamefont
	{Xiao}, \citenamefont {Bunch},\ and\ \citenamefont {Dunn}}]{boddeti13}%
\BibitemOpen
\bibfield  {author} {\bibinfo {author} {\bibfnamefont {N.~G.}\ \bibnamefont
		{Boddeti}}, \bibinfo {author} {\bibfnamefont {S.~P.}\ \bibnamefont {Koenig}},
	\bibinfo {author} {\bibfnamefont {R.}~\bibnamefont {Long}}, \bibinfo {author}
	{\bibfnamefont {J.}~\bibnamefont {Xiao}}, \bibinfo {author} {\bibfnamefont
		{J.~S.}\ \bibnamefont {Bunch}}, \ and\ \bibinfo {author} {\bibfnamefont
		{M.~L.}\ \bibnamefont {Dunn}},\ }\href@noop {} {\bibfield  {journal}
	{\bibinfo  {journal} {Journal of Applied Mechanics}\ }\textbf {\bibinfo
		{volume} {80}},\ \bibinfo {pages} {040909} (\bibinfo {year}
	{2013})}\BibitemShut {NoStop}%
\bibitem [{\citenamefont {Castellanos-Gomez}\ \emph {et~al.}(2015)\citenamefont
	{Castellanos-Gomez}, \citenamefont {Singh}, \citenamefont {van~der Zant},\
	and\ \citenamefont {Steele}}]{castellanos15_review}%
\BibitemOpen
\bibfield  {author} {\bibinfo {author} {\bibfnamefont {A.}~\bibnamefont
		{Castellanos-Gomez}}, \bibinfo {author} {\bibfnamefont {V.}~\bibnamefont
		{Singh}}, \bibinfo {author} {\bibfnamefont {H.~S.~J.}\ \bibnamefont {van~der
			Zant}}, \ and\ \bibinfo {author} {\bibfnamefont {G.~A.}\ \bibnamefont
		{Steele}},\ }\href {\doibase 10.1002/andp.201400153} {\bibfield  {journal}
	{\bibinfo  {journal} {Annalen der Physik}\ }\textbf {\bibinfo {volume}
		{527}},\ \bibinfo {pages} {27} (\bibinfo {year} {2015})}\BibitemShut
{NoStop}%
\bibitem [{\citenamefont {Isacsson}\ \emph {et~al.}(2016)\citenamefont
	{Isacsson}, \citenamefont {Cummings}, \citenamefont {Colombo}, \citenamefont
	{Colombo}, \citenamefont {Kinaret},\ and\ \citenamefont
	{Roche}}]{isacsson2017_review}%
\BibitemOpen
\bibfield  {author} {\bibinfo {author} {\bibfnamefont {A.}~\bibnamefont
		{Isacsson}}, \bibinfo {author} {\bibfnamefont {A.~W.}\ \bibnamefont
		{Cummings}}, \bibinfo {author} {\bibfnamefont {L.}~\bibnamefont {Colombo}},
	\bibinfo {author} {\bibfnamefont {L.}~\bibnamefont {Colombo}}, \bibinfo
	{author} {\bibfnamefont {J.~M.}\ \bibnamefont {Kinaret}}, \ and\ \bibinfo
	{author} {\bibfnamefont {S.}~\bibnamefont {Roche}},\ }\href@noop {}
{\bibfield  {journal} {\bibinfo  {journal} {2D Materials}\ }\textbf {\bibinfo
		{volume} {4}},\ \bibinfo {pages} {012002} (\bibinfo {year}
	{2016})}\BibitemShut {NoStop}%
\bibitem [{\citenamefont {Barton}\ \emph {et~al.}(2012)\citenamefont {Barton},
	\citenamefont {Storch}, \citenamefont {Adiga}, \citenamefont {Sakakibara},
	\citenamefont {Cipriany}, \citenamefont {Ilic}, \citenamefont {Wang},
	\citenamefont {Ong}, \citenamefont {McEuen}, \citenamefont {Parpia},\ and\
	\citenamefont {Craighead}}]{barton12}%
\BibitemOpen
\bibfield  {author} {\bibinfo {author} {\bibfnamefont {R.~A.}\ \bibnamefont
		{Barton}}, \bibinfo {author} {\bibfnamefont {I.~R.}\ \bibnamefont {Storch}},
	\bibinfo {author} {\bibfnamefont {V.~P.}\ \bibnamefont {Adiga}}, \bibinfo
	{author} {\bibfnamefont {R.}~\bibnamefont {Sakakibara}}, \bibinfo {author}
	{\bibfnamefont {B.~R.}\ \bibnamefont {Cipriany}}, \bibinfo {author}
	{\bibfnamefont {B.}~\bibnamefont {Ilic}}, \bibinfo {author} {\bibfnamefont
		{S.~P.}\ \bibnamefont {Wang}}, \bibinfo {author} {\bibfnamefont
		{P.}~\bibnamefont {Ong}}, \bibinfo {author} {\bibfnamefont {P.~L.}\
		\bibnamefont {McEuen}}, \bibinfo {author} {\bibfnamefont {J.~M.}\
		\bibnamefont {Parpia}}, \ and\ \bibinfo {author} {\bibfnamefont {H.~G.}\
		\bibnamefont {Craighead}},\ }\href {\doibase 10.1021/nl302036x} {\bibfield
	{journal} {\bibinfo  {journal} {Nano Letters}\ }\textbf {\bibinfo {volume}
		{12}},\ \bibinfo {pages} {4681} (\bibinfo {year} {2012})}\BibitemShut
{NoStop}%
\bibitem [{\citenamefont {Lifshitz}\ and\ \citenamefont
	{Cross}(2008)}]{lifshitz08}%
\BibitemOpen
\bibfield  {author} {\bibinfo {author} {\bibfnamefont {R.}~\bibnamefont
		{Lifshitz}}\ and\ \bibinfo {author} {\bibfnamefont {M.}~\bibnamefont
		{Cross}},\ }\href@noop {} {\bibfield  {journal} {\bibinfo  {journal} {Review
			of nonlinear dynamics and complexity}\ }\textbf {\bibinfo {volume} {1}},\
	\bibinfo {pages} {1} (\bibinfo {year} {2008})}\BibitemShut {NoStop}%
\bibitem [{\citenamefont {Hauer}\ \emph {et~al.}(2013)\citenamefont {Hauer},
	\citenamefont {Doolin}, \citenamefont {Beach},\ and\ \citenamefont
	{Davis}}]{hauer13}%
\BibitemOpen
\bibfield  {author} {\bibinfo {author} {\bibfnamefont {B.}~\bibnamefont
		{Hauer}}, \bibinfo {author} {\bibfnamefont {C.}~\bibnamefont {Doolin}},
	\bibinfo {author} {\bibfnamefont {K.}~\bibnamefont {Beach}}, \ and\ \bibinfo
	{author} {\bibfnamefont {J.}~\bibnamefont {Davis}},\ }\href@noop {}
{\bibfield  {journal} {\bibinfo  {journal} {Annals of Physics}\ }\textbf
	{\bibinfo {volume} {339}},\ \bibinfo {pages} {181} (\bibinfo {year}
	{2013})}\BibitemShut {NoStop}%
\end{thebibliography}

\begin{thebibliography}{6}%
	\makeatletter
	\providecommand \@ifxundefined [1]{%
		\@ifx{#1\undefined}
	}%
	\providecommand \@ifnum [1]{%
		\ifnum #1\expandafter \@firstoftwo
		\else \expandafter \@secondoftwo
		\fi
	}%
	\providecommand \@ifx [1]{%
		\ifx #1\expandafter \@firstoftwo
		\else \expandafter \@secondoftwo
		\fi
	}%
	\providecommand \natexlab [1]{#1}%
	\providecommand \enquote  [1]{``#1''}%
	\providecommand \bibnamefont  [1]{#1}%
	\providecommand \bibfnamefont [1]{#1}%
	\providecommand \citenamefont [1]{#1}%
	\providecommand \href@noop [0]{\@secondoftwo}%
	\providecommand \href [0]{\begingroup \@sanitize@url \@href}%
	\providecommand \@href[1]{\@@startlink{#1}\@@href}%
	\providecommand \@@href[1]{\endgroup#1\@@endlink}%
	\providecommand \@sanitize@url [0]{\catcode `\\12\catcode `\$12\catcode
		`\&12\catcode `\#12\catcode `\^12\catcode `\_12\catcode `\%12\relax}%
	\providecommand \@@startlink[1]{}%
	\providecommand \@@endlink[0]{}%
	\providecommand \url  [0]{\begingroup\@sanitize@url \@url }%
	\providecommand \@url [1]{\endgroup\@href {#1}{\urlprefix }}%
	\providecommand \urlprefix  [0]{URL }%
	\providecommand \Eprint [0]{\href }%
	\providecommand \doibase [0]{http://dx.doi.org/}%
	\providecommand \selectlanguage [0]{\@gobble}%
	\providecommand \bibinfo  [0]{\@secondoftwo}%
	\providecommand \bibfield  [0]{\@secondoftwo}%
	\providecommand \translation [1]{[#1]}%
	\providecommand \BibitemOpen [0]{}%
	\providecommand \bibitemStop [0]{}%
	\providecommand \bibitemNoStop [0]{.\EOS\space}%
	\providecommand \EOS [0]{\spacefactor3000\relax}%
	\providecommand \BibitemShut  [1]{\csname bibitem#1\endcsname}%
	\let\auto@bib@innerbib\@empty
	%</preamble>
	\bibitem [{\citenamefont {Amabili}(2008)}]{amabili08}%
	\BibitemOpen
	\bibfield  {author} {\bibinfo {author} {\bibfnamefont {Marco}\ \bibnamefont
			{Amabili}},\ }\href@noop {} {\emph {\bibinfo {title} {Nonlinear vibrations
				and stability of shells and plates}}}\ (\bibinfo  {publisher} {Cambridge
		University Press},\ \bibinfo {year} {2008})\BibitemShut {NoStop}%
	\bibitem [{\citenamefont {Timoshenko}\ and\ \citenamefont
		{Woinowsky-Krieger}(1959)}]{timoshenko59}%
	\BibitemOpen
	\bibfield  {author} {\bibinfo {author} {\bibfnamefont {Stephen}\ \bibnamefont
			{Timoshenko}}\ and\ \bibinfo {author} {\bibfnamefont {Sergius}\ \bibnamefont
			{Woinowsky-Krieger}},\ }\bibfield  {title} {\enquote {\bibinfo {title}
			{Theory of plates and shells},}\ }\href@noop {} {\  (\bibinfo {year}
		{1959})}\BibitemShut {NoStop}%
	\bibitem [{\citenamefont {Amabili}\ and\ \citenamefont
		{Breslavsky}(2015)}]{amabili15}%
	\BibitemOpen
	\bibfield  {author} {\bibinfo {author} {\bibfnamefont {Marco}\ \bibnamefont
			{Amabili}}\ and\ \bibinfo {author} {\bibfnamefont {Ivan~D}\ \bibnamefont
			{Breslavsky}},\ }\bibfield  {title} {\enquote {\bibinfo {title} {Displacement
				dependent pressure load for finite deflection of doubly-curved thick shells
				and plates},}\ }\href@noop {} {\bibfield  {journal} {\bibinfo  {journal}
			{International Journal of Non-Linear Mechanics}\ }\textbf {\bibinfo {volume}
			{77}},\ \bibinfo {pages} {265--273} (\bibinfo {year} {2015})}\BibitemShut
	{NoStop}%
	\bibitem [{\citenamefont {Lee}\ \emph {et~al.}(2008)\citenamefont {Lee},
		\citenamefont {Wei}, \citenamefont {Kysar},\ and\ \citenamefont
		{Hone}}]{hone08elastic}%
	\BibitemOpen
	\bibfield  {author} {\bibinfo {author} {\bibfnamefont {Changgu}\ \bibnamefont
			{Lee}}, \bibinfo {author} {\bibfnamefont {Xiaoding}\ \bibnamefont {Wei}},
		\bibinfo {author} {\bibfnamefont {Jeffrey~W.}\ \bibnamefont {Kysar}}, \ and\
		\bibinfo {author} {\bibfnamefont {James}\ \bibnamefont {Hone}},\ }\bibfield
	{title} {\enquote {\bibinfo {title} {Measurement of the elastic properties
				and intrinsic strength of monolayer graphene},}\ }\href {\doibase
		10.1126/science.1157996} {\bibfield  {journal} {\bibinfo  {journal}
			{Science}\ }\textbf {\bibinfo {volume} {321}},\ \bibinfo {pages} {385--388}
		(\bibinfo {year} {2008})}\BibitemShut {NoStop}%
	\bibitem [{\citenamefont {Nayfeh}\ and\ \citenamefont {Mook}(2008)}]{nayfeh08}%
	\BibitemOpen
	\bibfield  {author} {\bibinfo {author} {\bibfnamefont {Ali~H}\ \bibnamefont
			{Nayfeh}}\ and\ \bibinfo {author} {\bibfnamefont {Dean~T}\ \bibnamefont
			{Mook}},\ }\href@noop {} {\emph {\bibinfo {title} {Nonlinear oscillations}}}\
	(\bibinfo  {publisher} {John Wiley \& Sons},\ \bibinfo {year}
	{2008})\BibitemShut {NoStop}%
	\bibitem [{\citenamefont {Amabili}\ \emph {et~al.}(2016)\citenamefont
		{Amabili}, \citenamefont {Alijani},\ and\ \citenamefont
		{Delannoy}}]{amabili16}%
	\BibitemOpen
	\bibfield  {author} {\bibinfo {author} {\bibfnamefont {Marco}\ \bibnamefont
			{Amabili}}, \bibinfo {author} {\bibfnamefont {Farbod}\ \bibnamefont
			{Alijani}}, \ and\ \bibinfo {author} {\bibfnamefont {Joachim}\ \bibnamefont
			{Delannoy}},\ }\bibfield  {title} {\enquote {\bibinfo {title} {Damping for
				large-amplitude vibrations of plates and curved panels, part 2:
				Identification and comparisons},}\ }\href@noop {} {\bibfield  {journal}
		{\bibinfo  {journal} {International Journal of Non-Linear Mechanics}\
		}\textbf {\bibinfo {volume} {85}},\ \bibinfo {pages} {226--240} (\bibinfo
		{year} {2016})}\BibitemShut {NoStop}%
\end{thebibliography}
\end{document}